\documentclass[superscriptaddress,twocolumn,amsmath,amssymb]{revtex4-2} %twocolumn %onecolumn

\usepackage{xcolor}
\usepackage{color,soul}% highlighting using "\hl{}"
\usepackage{graphicx}% Include figure files
\usepackage{dcolumn}% Align table columns on decimal point
\usepackage{bm}% bold math
\usepackage{siunitx}
\usepackage[colorlinks=true,linkcolor=blue,citecolor=blue,urlcolor=blue]{hyperref}

\newcommand{\citeasnoun}[1]{Ref.~\cite{#1}}

\newcommand{\Figref}[1]{Figure~\ref{fig:#1}}
\newcommand{\figref}[1]{Fig.~\ref{fig:#1}}

\newcommand{\Eqref}[1]{Equation~(\ref{eq:#1})}
\renewcommand{\eqref}[1]{Eq.~(\ref{eq:#1})}

\newcommand{\eqrefs}[2]{Eqs.~(\ref{eq:#1},\ref{eq:#2})}
\newcommand{\Eqrefrange}[2]{Equations~(\ref{eq:#1}--\ref{eq:#2})}

\newcommand{\secref}[1]{Sec.~\ref{sec:#1}}

\renewcommand{\AA}{\mathbb{A}}
\newcommand{\BB}{\mathbb{B}}

\newcommand{\XX}{\mathbb{X}}
\newcommand{\LL}{\mathbb{L}}

\newcommand{\SM}{\textrm{SM}}
\DeclareMathOperator{\Tr}{Tr}

\begin{document}

\title{Many Physical Design Problems are Sparse QCQPs}

\author{Shai Gertler}
\affiliation{Department of Applied Physics and Energy Sciences Institute, Yale University, New Haven, CT 06520, USA}
\author{Zeyu Kuang} 
\affiliation{Department of Applied Physics and Energy Sciences Institute, Yale University, New Haven, CT 06520, USA}
\author{Colin Christie}
\affiliation{Department of Applied Physics and Energy Sciences Institute, Yale University, New Haven, CT 06520, USA}
\author{Owen D. Miller}
\affiliation{Department of Applied Physics and Energy Sciences Institute, Yale University, New Haven, CT 06520, USA}

\begin{abstract}
Physical design refers to mathematical optimization of a desired objective (e.g. strong light--matter interactions, or complete quantum state transfer) subject to the governing dynamical equations, such as Maxwell's or Schrodinger's differential equations. Computing an optimal design is challenging: generically, these problems are highly nonconvex and finding global optima is NP hard. Here we show that for linear-differential-equation dynamics (as in linear electromagnetism, elasticity, quantum mechanics, etc.), the physical-design optimization problem can be transformed to a sparse-matrix, quadratically constrained quadratic program (QCQP). Sparse QCQPs can be tackled with convex optimization techniques (such as semidefinite programming) that have thrived for identifying global bounds and high-performance designs in other areas of science and engineering, but seemed inapplicable to the design problems of wave physics. We apply our formulation to prototypical photonic design problems, showing the possibility to compute fundamental limits for large-area metasurfaces, as well as the identification of designs approaching global optimality. Looking forward, our approach highlights the promise of developing bespoke algorithms tailored to specific physical design problems.
\end{abstract}
%State-of-the-art approaches revolve around ``inverse design'' or neural-network techniques that are quite general, but which exploit no mathematical structure beyond differentiability. 

\maketitle
Precise fabrication techniques and high-quality materials are enabling unprecedented control over photonic and quantum physical systems, opening exciting frontiers for applications from metasurface optics~\cite{Yu2014,Lalanne2017} to quantum computing~\cite{Heeres2017,Lam2021}. Yet this flexibility is accompanied by a proportionate complexity: the design spaces are enormous, and classical and quantum wave-interference effects make them highly oscillatory and nonconvex. The corresponding design problem, of discovering optimal patterns in space and/or time, is generically NP hard~\cite{Angeris2021}. In this article, we identify a surprising mathematical structure in design problems arising from the typical differential equations of physics: in many cases, they can be transformed to sparse, quadratically constrained quadratic programs (sparse QCQPs). It is well known that sparse QCQPs can be close to convex optimization problems, in a sense we make precise below, implying that the transformed design problems may have significantly smoother landscapes and that global optima may be easier to approach. The seemingly impenetrable nonconvexity of conventional formulations of physical design has limited scientists and engineers to black-box algorithms or gradient-based local optimization. Our transformations enable direct connection to a wide swath of modern optimization theory, where advances in convex optimization and semidefinite programming have enabled remarkable progress in trajectory optimization (e.g., rocket landings)~\cite{Acikmese2007,Sagliano2019}, matrix completion (recommendation systems)~\cite{Candes2010,Candes2012,Kalofolias2014}, compressed sensing (network routing)~\cite{Tropp2006,Tropp2010}, and more. We show that for prototypical problems in domains such as photonic design and quantum control, one can leverage the sparse-QCQP structure to find fundamental limits for large-scale devices, and offer the possibility of identifying designs near global optimality. We develop a general theory for connecting these ideas in nanophotonics, quantum control, and beyond, and offer a new, convex-optimization-based direction for physical design. 

%The design transformation arises by replacing differential-equation constraints with conservation laws, mimicking Kirchhoff's current and voltage laws but for the differential equations of Maxwell, Schrodinger, and others. 

%The transformed constraint set can be understood as the geometric intersection of ellipsoids and hyperboloids in a high-dimensional space, but ``lifting'' the problem to higher dimensions creates a set of linear constraints, the intersections of hyperplanes and half-spaces. Many of these ideas have been developed in various arenas of science and engineering, but were presumably inapplicable to the control and design of physical systems. 

In the 1990's it was recognized that a key dividing line between ``easy'' and ``hard'' optimization problems was not linearity versus nonlinearity, but rather convexity versus nonconvexity~\cite{Rockafellar1993}. Convex optimization problems, with convex objective functions and convex constraint sets, share a remarkable property: all local optima are global optima~\cite{Boyd2004}. Convex optimization problems can be solved efficiently (in polynomial time) for global optima, e.g., by now-standard interior-point methods~\cite{Boyd2004}. Original applications of convex optimization primarily arose in operations research, but over the past two decades, a surprising number of optimization problems have been shown to be transformable to convex, or nearly convex, formulations. In addition to the three applications mentioned above, further examples include ptychography~\cite{Horstmeyer2015,Yurtsever2021}, in which wide-field-of-view, high-resolution images are formed from low-resolution samples, and phase retrieval~\cite{Candes2013,Goldstein2018}, in which one infers phase values from intensity measurements. By contrast, quantum and classical wave dynamics create interference patterns and oscillatory objective landscapes that are clearly nonconvex.

%The sparse QCQPs that we identify below are not themselves convex, but they have known (convex) dual programs whose solutions are guaranteed to represent bounds for the problem of interest. Moreover, the only nonconvexity of the QCQPs arises from a rank constraint, and the sparse nature of differential operators can lead to guarantees of low-rank solutions within the dual program. 

Confronted by the nonconvexity of physical design, two approaches are typically taken. One is to compute gradients of the objective function with respect to the designable degrees of freedom and ``ascend'' or ``descend'' using the gradient. A particular breakthrough in the efficiency of such algorithms came from the ``adjoint variable'' technique~\cite{Strang2007,Temes1977,Jameson1988} (or ``backpropagation''~\cite{Werbos1990}), which forms the basis of approaches known as ``inverse design''~\cite{Miller2012,Piggott2015,Sell2017} and ``topology optimization''~\cite{Jensen2011,Lin2019,Christiansen2019} in photonics, and  ``GRAPE'' and ``Krotov'' in quantum control~\cite{Krotov1993,Khaneja2005,Brif2010,DeFouquieres2011,Goerz2019}. Such inverse-design techniques have been used across linear, nonlinear~\cite{Tortorelli1991,Hughes2018}, and even chaotic~\cite{Wang2013} physical systems. However, as gradient-based techniques, they may get stuck in low-quality local extrema, and they offer no insight into global bounds (or, ``fundamental limits'').

Alternatively, one can use techniques intended for global optimization; early interest in evolutionary algorithms has largely been supplanted by modern research in neural networks. Yet neural networks typically excel in scenarios where modeling is difficult and an extraordinary amount of data is availabile (such as image recognition and large language models). In physical design, data is expensive to generate (full differential-equation simulations), we have exact models (the differential equations themselves), and learning across architectures, material platforms, and frequency ranges has proven difficult. There have been a number of pioneering efforts in applying machine learning to physical design~\cite{Peurifoy2018,Jiang2019,Raissi2019,Kiarashinejad2020}, but it is highly unclear whether they can generally outperform gradient-based inverse design once all computational costs associated with training are included.

A complementary approach to bottom-up, iterative optimization has been a recent surge of interest in fundamental limits, particularly in optics and photonics~\cite{Yu2010,Miller2016,Miller2015,Hugonin2015,Miller2017,Sanders2018,Zhang2019,Shim2019,Molesky2020a,Gustafsson2020,Trivedi2020,Molesky2020b,Kuang2020b,Molesky2020,Presutti2020,Li2022,Miller2023}. Conventionally, bounds such as the Yablonovitch enhancement limit in electromagnetism~\cite{Yablonovitch1982}, or the Mandelstamm--Tamm bounds in quantum control~\cite{Mandelstam1945}, arise from identifying singular constraints in physical systems (density of states and energy--time uncertainty, for the two examples). Recently, an alternative approach was proposed in which computational bounds could be found from an infinite set of constraints implied by integral~\cite{Kuang2020,Molesky2020,Zhang2021} or differential~\cite{Angeris2021,angeris2022bounds} equations. The differential-equation-based approaches requires a transformation to real-valued variables and a relaxation of the design problem (cf. \SM), while the matrices in the integral QCQPs are dense, due to their integral-equation origins, leading to two issues: First, solving their semidefinite relaxations for bounds requires $O(N^4)$ computational time, for $N$ degrees of freedom, which quickly becomes prohibitive and restricts analysis to few-degree-of-freedom systems. Second, the ranks of the solutions of the semidefinite relaxations are unbounded; as we describe in depth below, this means that the solutions of the semidefinite programs may have little-to-no correlation with physically meaningful structures, let alone optimal designs.

The approach we describe here is a new, joint formulation of physical design and its fundamental limits. Starting with a generic specification of a physical design problem with differential-equation constraints, we describe general techniques for transformation to a sparse QCQP optimization problem. We show that the new problem is mathematically equivalent to the original, yet with a structure more similar to that of ptychography, trajectory optimization, or matrix completion, instead of conventional physical design. The ``dual''~\cite{Boyd2004} of our sparse-QCQP problem is a (convex) semidefinite program (SDP),  whose solution always represents a fundamental limit across the design space of interest. Moreover, the sparse nature of the differential operators leads to bounds on the rank of the solution of the SDP, which often must be small (e.g., independent of the diameter of a metasurface). A rank-one solution would represent the globally optimal design across the design space; low-rank solutions can potentially be regularized towards rank-one solutions. We give an example of exactly such an approach, leading to a plausibly globally optimal nanophotonic structure for a prototype problem on which ``inverse design'' struggles to overcome moderate-quality local optima. Our formulation enables the identification of bounds for optimal design with linear differential equations, and offers a new approach to circumvent the oscillatory landscapes that plague the design of wave-based systems.

\section{Reformulating Physical Design}
The prototypical design problem has a physical field $\psi$ (electromagnetic, quantum, etc.), designable degrees of freedom  $\chi$ (a susceptibility, a control-Hamiltonian amplitude, etc.), and a differential equation linking the two. We consider differential equations that are linear in the field variable. We denote by $\xi$ a vector-field excitation (which can be zero for an eigenproblem), representing an incoming wave or initial condition. The goal is to maximize (or minimize) some objective $f(\psi)$ that is a function of the field variable. Then the design problem can be written:
\begin{equation}
    \begin{aligned}
    & \underset{\chi,\psi}{\text{max.}}
    & & f(\psi) \\
    & \text{s.t.}
    & & \mathbb{L}(\chi) \psi - \xi = 0,
    \end{aligned}
    \label{eq:opt_prob1}
\end{equation}
where $\LL$ represents the linear differential operator. We include $\chi$ as an argument to $\LL$ to emphasize the dependence of $\LL$ on the design variables. In nanophotonics, where one might seek an optimal susceptibility pattern $\chi(x)$ in space to harness electromagnetic waves of frequency $\omega$, the differential operator is $\LL = \nabla \times \nabla \times - \left[1 + \chi(x)\right] \omega^2$. In quantum control, one might seek an optimal pulse sequence $\chi(t)$ representing the time-dependent amplitude of a control Hamiltonian, $H_c(t)$, relative to a background Hamiltonian, $H_0(t)$, in the Schrodinger operator, $\LL = i \frac{d}{dt} - \left[H_0(t) + \chi(t) H_c(t) \right]$. And similarly for elastic waves, with the Christoffel equation, and many other linear differential equations of physics. To simplify the exposition and clarify the complexity analysis, we will assume any sufficiently high-resolution numerical discretization into $N$ spatial/temporal/polarization/Hilbert-space/etc. degrees of freedom, in which case the variables $\psi$ and $\chi$ are $N\times1$ vectors and $\LL$ is an $N \times N$ matrix (with suitable boundary conditions encoded). The task of scientists and engineers in disciplines from aerodynamic wing design to nanophotonics to quantum information theory is to solve optimization problems with forms similar to \eqref{opt_prob1}.

\subsection{Physical design problems as QCQPs}
\Eqref{opt_prob1} is computationally prohibitive to solve for a global optimum because the differential-equation constraint is nonconvex in the variables $\psi$ and $\chi$. (It is ``bilinear'' in the pair of variables $\psi$ and $\chi$, but that is of little help in searching for global optima.) In this section, we show how to convert the differential-equation constraints to quadratic constraints, ultimately leading to the sparse-QCQP formulation of the design problem.

We consider the prototypical case in which there is a binary choice for the design variable over each designable DOF, denoted $\chi_1$ and $\chi_2$, such as, for example, designing a pattern of air holes in silicon, or a sequence of on/off pulses in quantum control. We further assume, for simplicity, that our design space has the full $N$ degrees of freedom shared by the response fields. Typically the number of degrees of freedom will be smaller than $N$; for example, perfectly matched layers are not part of the designable region. The constraints associated with nondesignable regions are straightforward and described in detail in the {\SM}, and do not change the analysis below.  Our approach generalizes to different numbers of degrees of freedom as well as to multilevel design problems, as we discuss in the {\SM}, but again these assumptions will clarify the key logic and analysis. It is simple to enumerate the design space: there are $N$ degrees of freedom, each of which can take one of two values, such that there are $2^{N}$ possible designs. The $2^N$ possible designs each have fields $\psi$ that must satisfy the $N$ differential-equation constraints of \eqref{opt_prob1}.  

The crucial insight is as follows: at each spatial/temporal point $i$ in the design space, either $\mathbb{L}(\chi_1) \psi - \xi\big\rvert_i = 0$ or $\mathbb{L}(\chi_2) \psi - \xi\big\rvert_i = 0$. (The ``$\big\rvert_i$'' symbol should be interpreted, for example, as selecting a single row of a finite-difference matrix, or as integrating against a single basis function in a finite element discretization.) Consider the scalar-field case, for which each expression $\LL(\chi_{1,2}) \psi - \xi\big\rvert_i$ is a scalar expression. Then, enforcing a logical OR condition between two scalar values is exactly equivalent to forming a \emph{single} constraint in which the two scalar expressions are multiplied together; in other words, enforcing $(a==0)\textrm{ OR }(b==0)$ is equivalent to enforcing $ab==0$. For the general vector-field case, the vector OR condition can be replaced by pairwise inner-product constraints. Each of these inner products is of the form:
\begin{align}
    \left[\mathbb{L}(\chi_1)\psi - \xi\right]^\dagger \mathbb{D}_i  \left[\mathbb{L}(\chi_2)\psi - \xi\right] =0,
    \label{eq:const1}
\end{align}
where $\mathbb{D}_i$ is the matrix that encodes the evaluation of the two terms in square brackets associated with a particular susceptibility degree of freedom and $\dagger$ represents the adjoint (conjugate transpose) operation. For example, in a finite-difference scheme, $D_i$ is a block-diagonal matrix (or any linear combination thereof) whose only nonzero elements occur at point $i$ in space/time, with nonzero off-diagonal elements only in polarization/Hilbert space. Enforcing \eqref{const1} over all independent block-diagonal $\mathbb{D}_i$ matrices is equivalent to enforcing the logical-OR condition at all points in space/time, as intuitively justified above and rigorously proven in the {\SM}. The number of independent constraints of the form of \eqref{const1} scales linearly with $N$, as one intuitively expects for $N$ degrees of freedom.

Clearly the constraints of \eqref{const1} are necessarily satisfied by any solution of the original design problem, \eqref{opt_prob1}. Perhaps surprisingly, the constraints of \eqref{const1} are also \emph{sufficient} conditions for the design problem. Hence we can discard the original differential equation constraints, and specify a new, equivalent design problem:
\begin{equation}
    \begin{aligned}
    & \underset{\psi}{\text{max.}}
    & & f(\psi) \\
    & \text{s.t.}
    & & \left[\mathbb{L}(\chi_1)\psi - \xi\right]^\dagger \mathbb{D}_i  \left[\mathbb{L}(\chi_2)\psi - \xi\right] =0 \quad \forall i.
    \end{aligned}
    \label{eq:opt_prob2}
\end{equation}
To see why the constraints of \eqref{opt_prob2} are sufficient conditions, consider an optimal solution $\psi$ of \eqref{opt_prob2}. From $\psi$, one can check every constraint $i$ to determine at each space/time point whether $\chi_1$ or $\chi_2$ is the optimal design variable at that point. The resulting pair of $\psi$ and $\chi$ will then necessarily satisfy the constraint of \eqref{opt_prob1}, implying that any solution of \eqref{opt_prob2} is a viable solution of \eqref{opt_prob1}. As the optimal solutions of each problem are within the other's feasible set, and the objectives are the same, the optimal solutions must coincide. Hence, the two design problems have equivalent optima.

What is the intuition behind the quadratic constraints of \eqref{opt_prob2}? They turn out to have a simple, general physical interpretation. Each constraint in \eqref{opt_prob2} represents a complex-valued \emph{conservation law} at each point in space/time of the designable domain (cf. {\SM} for a derivation). In electromagnetism, they represent local conservation of real and reactive power flow. In quantum control, they represent local conservation of complex probability. And similarly for other physical systems. A close analog arises in circuit design. When one wants to find the steady-state response of a complex configuration of resistors, capacitors, and inductors, the first choice typically is not solving circuit differential equations. Instead, one solves Kirchhoff's current and voltage laws, two conservation laws for every node of the system. The constraints of \eqref{opt_prob2} are akin to generalized Kirchhoff's laws for any linear-differential-equation-based design problem.
\begin{figure*}[htb]
    \centering
    \includegraphics[width=.85\textwidth]{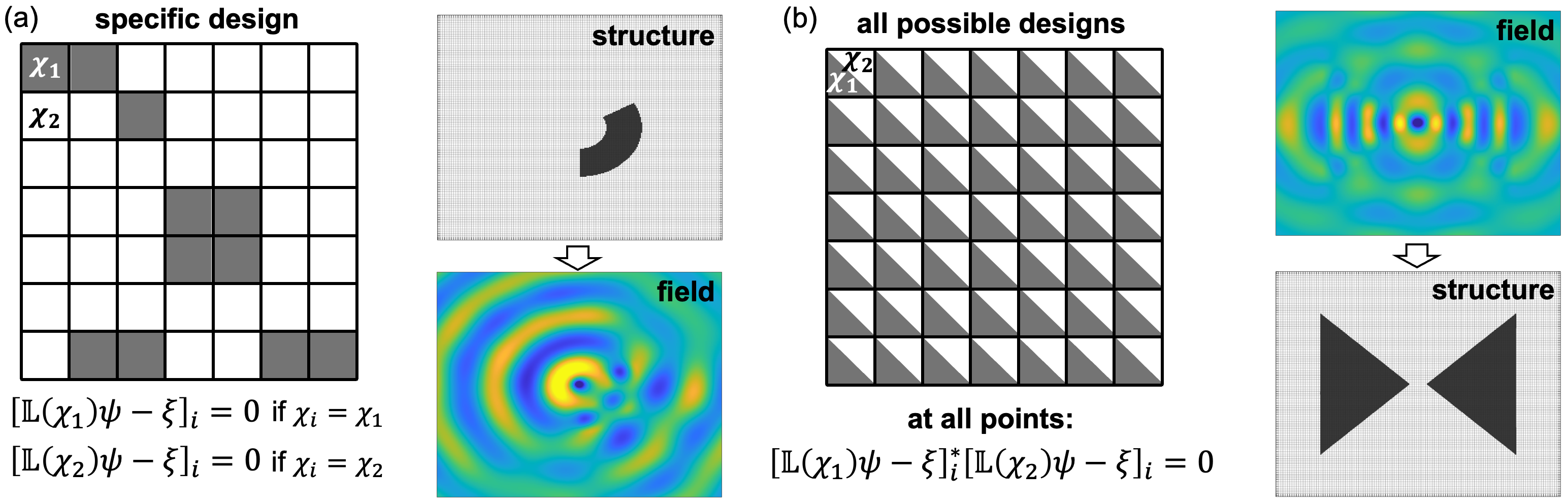}
            \caption{(a) The usual writing of a differential equation specifies different operator values at different space/time points according to the design variable (e.g. material susceptibility) $\chi$. One can specify a structure (top right, a material with $\chi=3$ and approximately half-wavelength thickness, for a finite-difference discretization of the Helmholtz equation) and compute the corresponding field response (bottom right). (b) Our approach identifies quadratic constraints satisfied by all possible designs. Any field that satisfies these constraints (e.g. top right) will imply a corresponding design (bottom right, $\chi=3$) that produces the field. If the field optimizes an objective of interest, then the corresponding structure will be an optimal design.}
    \label{fig:FTS}
\end{figure*}

The design problem of \eqref{opt_prob2} has a different mathematical structure from \eqref{opt_prob1}. First, in \eqref{opt_prob2}, the design variable $\chi$ is no longer a degree of freedom. Instead, the $2^N$ possible design configurations are encoded in the $O(N)$ quadratic constraints of \eqref{opt_prob2}. Now, the only degrees of freedom are in the field variable $\psi$, and the constraints are all quadratic forms in $\psi$. Objective functions $f(\psi)$ of interest, such as power flow or momentum transfer in electromagnetism or state transfer, fidelity, and related objectives in quantum control, will also be quadratic (or linear) forms of $\psi$. Hence, our new design problem is a \emph{quadratically constrained quadratic program}, or QCQP.

\Figref{FTS} schematically illustrates the change in viewpoint brought on by the transformation to a QCQP. We consider as an example the Helmholtz equation, in which wave speed can take values 1 (background) or 1/2 (patterned material), which corresponds to susceptibility values of $\chi_2 = 1$ and $\chi_1 = 3$, respectively. The domains depicted in \figref{FTS} are 5 wavelengths in size (with half-wavelength perfectly matched layers surrounding them). \Figref{FTS}(a) shows the specification of a structure in space, which prescribes the precise form of the corresponding linear operator. From the structure-specific operator, one can compute the fields. This is a powerful simulation tool, but it inhibits optimization approaches beyond gradient descent (local structural perturbations). \Figref{FTS}(b) shows the QCQP perspective, in which quadratic constraints are formed that must be satisfied by all possible designs. Optimizing over fields that satisfy these constraints, for example via semidefinite-programming techniques, can lead to discovery of optimal designs.

\subsection{Many sparse QCQPs are nearly convex}
The next question is whether the QCQP of \eqref{opt_prob2} offers mathematical advantages over the original problem. QCQPs are generically NP hard~\cite{Sahni1974,Park2017}, just as the original problem is. (If this were not true, \eqref{opt_prob2} would prove that P = NP.) Yet QCQPs arise across many areas of science and engineering, and there is well-developed machinery for exploiting their specific mathematical structure~\cite{Boyd1994,Goemans1995,Luo2010,Sojoudi2012,Candes2013,Horstmeyer2015}. Geometrically, a set of quadratic constraints represents the intersection points of hyperboloids and ellipsoids; direct optimization over such intersections requires combinatorial algorithms. However, one can ``lift'' the problem to a higher-dimensional space, where the quadratic constraints in the original variables become \emph{linear} (and therefore convex) constraints in a new matrix variable. This process introduces two additional constraints: a (convex) positive-semidefinite matrix constraint, and a nonconvex constraint on the rank of the new matrix. In this new, higher-dimensional setting, the objective and all constraints are convex, except for the rank constraint. In essence, all of the nonconvexity of the problem is isolated into the single rank constraint. This offers two exciting possibilities. First, one can simply drop the rank constraint, leaving a convex problem whose efficiently computable solution will always represent a bound, or fundamental limit, on the problem of interest. Second, the sparsity of differential operators can in some cases provably lead to low-rank solutions of this relaxed, convex problem, suggesting that many physical design problems may be ``nearly convex'' in this high-dimensional space. Then one may be able to introduce a rank-penalizing regularization that leads to high-quality, high-performance optimal designs that are superior to those found by typical gradient-descent-based techniques. In this section we first introduce the general ``lifting'' procedure, then we describe our proposed convexity-related approach to fundamental limits and novel designs.

The QCQP of \eqref{opt_prob2} can be expressed generally as:
\begin{align}
\begin{split}
    \max_{\psi} \quad\ &\psi^\dagger \mathbb{A} \psi + \text{Re}\left[\alpha^\dagger \psi \right]\\ 
    \text{such that}\quad\ &\psi^\dagger \mathbb{B}_i \psi + \text{Re}\left[\beta_i^\dagger \psi \right] + c_i=0,
\label{eq:QCQP}
\end{split}
\end{align}
where $i$ iterates over all constraints (space/time/polarization/etc.), $\AA$ and $\alpha$ are defined by the objective of interest, and equations are kept real-valued by separating the real and imaginary parts of the constraints of \eqref{opt_prob2}. ``Lifting'' is now well-known and well-understood (\citeasnoun{Luo2010} is one review); we include here a brief summary. The objective and each constraint of \eqref{QCQP} are scalars, and one can trivially take the matrix trace of each. A trace is invariant under cyclic permutations; for example, $\Tr(\psi^\dagger \AA \psi) = \Tr(\AA \psi\psi^\dagger)$. If we denote a new matrix variable $\XX = \psi\psi^\dagger$, the quadratic form $\psi^\dagger \AA \psi$ becomes a linear form, $\Tr(\AA\XX)$, in the new, higher-dimensional matrix variable $\XX$. The linear forms in the original equation are straightforward to handle with conventional methods~\cite{Luo2010}, ultimately leading to modified matrices $\AA$ and $\BB_i$ whose expressions are explicitly given in the {\SM}. However, the entries of the matrix variable $\XX$ are not free to take on any values; because $\XX$ represents the outer product of a vector with itself, it must have rank 1 and be positive semidefinite. Following these steps, we can rewrite the QCQP of \eqref{opt_prob2} and \eqref{QCQP} as:
\begin{align}
    \max_{\XX} \quad &\Tr(\AA \XX) \label{eq:QCQP2} \\
    \text{such that}\quad & \Tr(\BB_i \XX) = b_i \nonumber \\
    &  \XX \geq 0 \nonumber \\
    &  \operatorname{rank}(\XX) = 1.  \nonumber
\end{align}
The trace-based objective and constraints are linear in $\XX$, while the positive-semidefinite requirement is a convex cone constraint. Whereas each differential-equation-based constraint of the original problem was nonconvex, our new, equivalent problem in \eqref{QCQP2} has a convex objective and many convex constraints; the only nonconvex part of the problem is the rank-one constraint.

The observation that the rank-1 constraint is the only nonconvex part of the problem offers insights into ``how convex'' the design problem is in the lifted, higher-dimensional space. If we just drop the rank-1 constraint, the resulting problem, a semidefinite program (SDP), is convex. The question, then, is how large of a penalty one pays by dropping the rank-1 constraint and solving the SDP. For the famous NP-hard ``max-cut'' problem, in which one seeks to identify the maximum number of edges in a graph partition, an SDP relaxation is guaranteed to have a solution that is within 12\% of the global optimum~\cite{Goemans1995}. For most examples of lifting there is no such known bound; instead, the rank of the solution of the SDP is used as a measure of how convex the problem is. For example, if the rank of the SDP solution is quite large, then the solutions of the SDP and the original QCQP may be wildly different. In that case, the rank-1 constraint is clearly important for finding the correct solution, and it is unclear if a convex problem can lead one close to the optimal solution. By contrast, if the rank of the convex problem is small, then one can imagine making slight tweaks to the problem (e.g. adding a regularization or penalty function) to find a rank-1 solution of a related convex problem, which may be a high-quality solution of the original problem. (We show a numerical example of this type below.) Hence, the rank of the relaxed problem, the SDP, is an indication of how nonconvex the design problem is.

Strikingly, in certain physical scenarios we can use the sparsity of differential operators to \emph{prove} that the rank of the computed solution of the relaxed problem is ``small;'' in particular, that the rank of the solution is bounded above by a constant that is independent of the long dimension of the problem. Bounds on the solutions of SDPs are common in problems with sparse matrices~\cite{Vandenberghe2015}, which we use in tandem with the differential operators that comprise $\AA$ and $\BB_i$ of \eqref{QCQP2}. The matrix entries of the variable $\XX$ form a set of vertices, and the nonzero entries of the objective and data matrices, $\AA$ and $\BB_i$, comprise edges of an undirected graph on those vertices. If the graph is chordal, then the size of the largest clique is known to be a bound on the maximum rank of a solution of the SDP~\cite{Vandenberghe2015}. (A brief review of the relevant graph-theoretic terms is provided in the {\SM}.) If the graph is not chordal, then a bound can be found from any chordal extension of the graph. As we show in the SM, for any linear-differential-equation-based physical design problem that has one ``long'' dimension (e.g. waveguides, metasurfaces with translational or rotational symmetry, quantum control problems with time as the long dimension), the clique number is bounded above by a problem-size-independent constant. These physical design problems are provably ``close'' to convex optimization problems.

\begin{figure*}[tb]
    \centering
    \includegraphics[width=.85\textwidth]{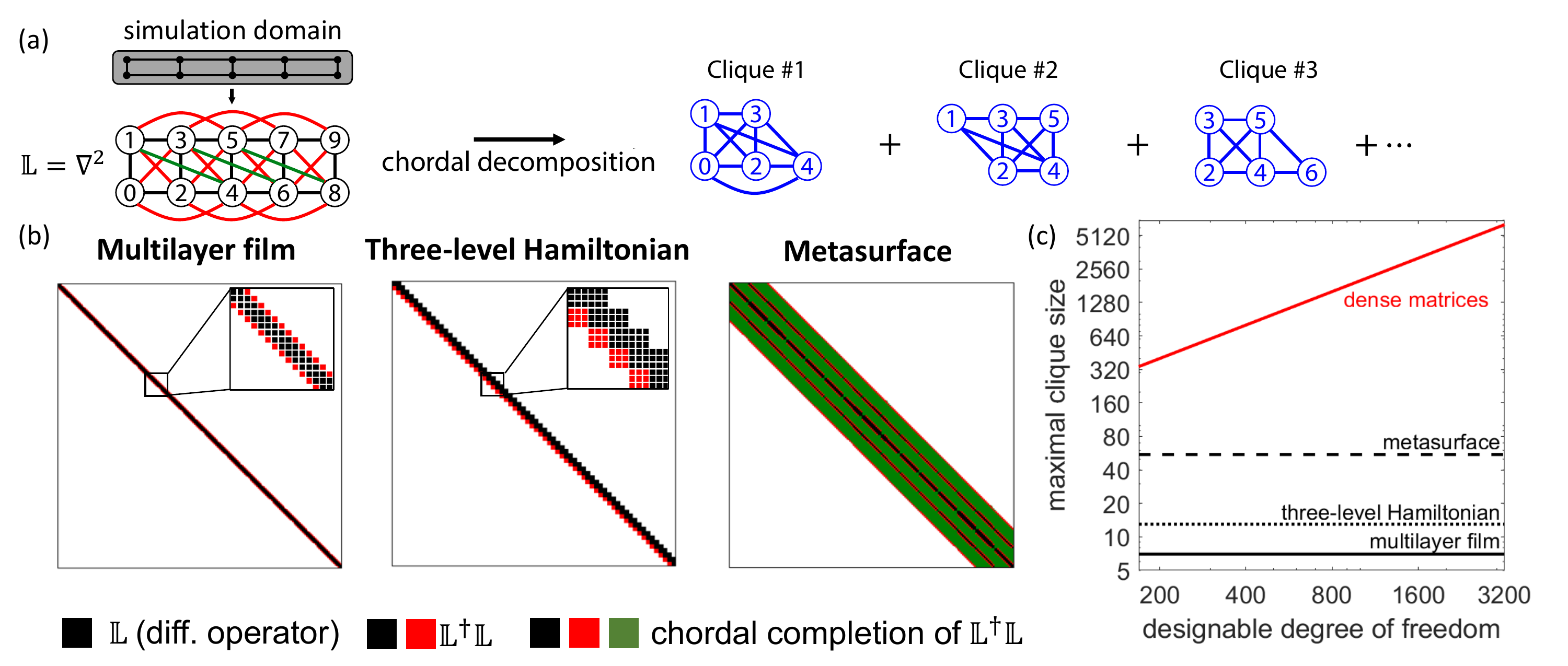}
    \caption{(a) Differential equations on any domain (black grid) have a sparsity pattern that can be represented in an undirected graph (vertices 0--8). The chordal completion of that graph can be decomposed into ``cliques,'' smaller domains on which the problem can be defined. (b) The rank of the corresponding SDP, a measure of the nonconvexity of the design problem, is bounded above by the largest clique size. For dense matrices (as arise in integral equations), this is simply proportional to the size of the problem (red). By contrast, the sparsity of differential operators leads to maximum clique sizes that are bounded above by a constant unrelated to the ``long'' dimension of a physical-design problem. (C) Chordal completions of the sparsity patterns arising in various differential-equation-based design problems in electromagnetism and quantum information theory.}
    \label{fig:sparsity}
\end{figure*}
\subsection{A convex optimization paradigm for physical design}
\label{sec:convex}
In this section, we formulate two convex-optimization-related approaches to the design problem of \eqref{QCQP2}. First, we define the SDP discussed above, with the rank-1 constraint dropped from the problem, which leads to bounds on any design problem of interest. Our second approach is to augment the SDP with a penalty term (typically convex) that encourages very-low-rank or even rank-1 solutions to the semidefinite program. This offers a new approach to designing physical structures, which comes with the possibility of avoiding many low-quality local optima.

Our first formulation simply drops the rank-1 constraint from \eqref{QCQP2}, leaving the SDP:
\begin{align}
    \max_{\XX} \quad &\Tr(\AA \XX) \label{eq:SDP} \\
    \text{such that}\quad & \Tr(\BB_i \XX) = b_i \nonumber \\
    &  \XX \geq 0. \nonumber
\end{align}
\Eqref{SDP} is a relaxation of the original physical design problem, and as a convex optimization problem, its global optima can be found efficiently via interior-point methods~\cite{Boyd2004}. A key feature of the solution is that it will always represent an upper (lower) bound to the objective to be maximized (minimized). This results directly from the dropping of the rank-one constraint; alternatively, it is known that \eqref{SDP} is equivalent to the dual problem of \eqref{QCQP2}, and dual-problem solutions always bound their ``primal'' counterparts. A key feature of \eqref{SDP} is that it can be decomposed into smaller, coupled sub-problems when the matrices $\AA$ and $\BB_i$ are sparse. From \eqref{opt_prob2}, one can see that the sparsity of the matrices $\BB_i$ is determined primarily by $\LL_i$, the differential operator, which is typically (e.g. in the case of the Laplacian) quite sparse. \figref{sparsity} demonstrates the sparsity patterns of the product $\LL^\dagger \LL$ that arises in \eqref{SDP}, for the differential operators of electromagnetism and quantum mechanics. Significant dimensionality reduction requires a strong form of sparsity: not only many zero entries, but also \emph{chordal sparsity}. The chordal completion of the $\LL^\dagger \LL$ operator is depicted in green in \figref{sparsity}(b); the sparsity of this matrix determines the extent to which the SDP can be compressed (losslessly). Moreover, a sparse chordal completion leads to small maximal clique sizes, as depicted in \figref{sparsity}(c), in stark contrast to SDPs formed from dense matrices. These maximal clique sizes are bounded above by a constant unrelated to the long dimension of the problem, suggesting that the SDP solutions may be informative for many design problems, which we support with numerical evidence in the next section.
\begin{figure*}[tb]
    \centering
    \includegraphics[width=0.9\textwidth]{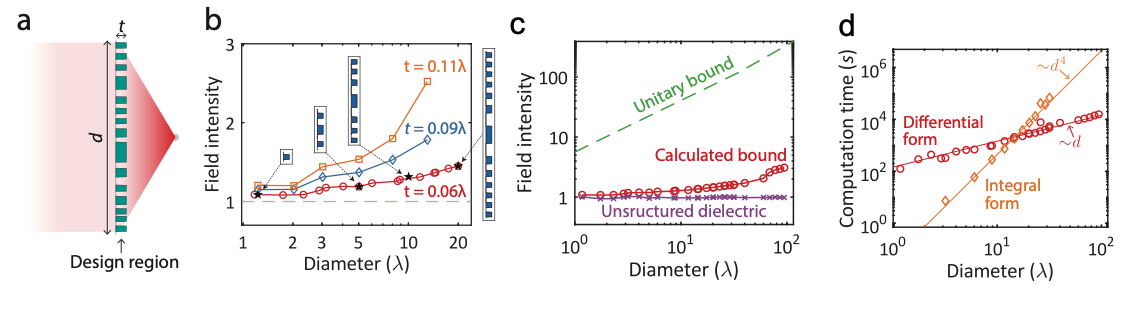}
    \caption{(a) Metalenses use nanophotonic patterning to focus incoming light. As a function of the diameter $d$ and thickness $t$ of the metalens, bounds on the maximal focal-point field intensity can be computed, shown in (b), using the SDP of \eqref{SDP}. If the solution is low rank, as for the red curve, the solution of the SDP can be used to find designs (field intensity in black markers, structure in insets) that nearly achieve the global bound. (c) Conservation-law bounds can be much smaller than simpler approaches, such as unitarity. (d) The differential form of the Maxwell equations in the SDP leads to much \emph{linear-in-diameter} scaling of computation times, enabling bounds for diameters up to 100 wavelenths.}
    \label{fig:metalens}
\end{figure*}

%In addition to the solution objective providing a global bound, the solution $\XX$ offers useful design information. A non-rank-one solution will not map exactly to a binary design solution of the original problem, \eqref{opt_prob1} or \eqref{opt_prob2}, but there are a variety of matrix-algebraic and randomized algorithms for identifying high-performance rank-one approximations that \emph{do} map to binary designs. Throughout the examples below we demonstrate examples of solving the convex program of \eqref{SDP}, instead of any of the nonconvex forms of \eqref{opt_prob1} through \eqref{QCQPs}, to identify fundamental limits and high-performance (wd?) designs in photonic and quantum settings.

The second approach is to augment the semidefinite program of \eqref{SDP} to promote solution matrices with ranks close to or equal to one. While the rank operator itself is a nonconvex quantity, there are well-known proxy quantities that promote low rank, such as the matrix trace, which is known to be the best convex lower bound on rank for matrices with constrained singular values~\cite{Fazel2002}. In feasibility problems, as occur, for example, in ptychographic imaging, one replaces the rank-1 constraint with an objective that promotes low rank~\cite{Horstmeyer2015}. For design problems, in which there is already an objective of interest, we can add a \emph{regularizer}, i.e., a penalty for high-rank solutions, to the original objective. If we denote by $\mathcal{R}(\XX)$ a functional proxy for the rank of a matrix $\XX$, then we can modify \eqref{SDP} to form the augmented program,
\begin{align}
    \max_{\XX} \quad &\Tr(\AA \XX) - \gamma \mathcal{R}(\XX) \label{eq:SDP2} \\
    \text{such that}\quad & \Tr(\BB_i \XX) = b_i \nonumber \\
    &  \XX \geq 0, \nonumber
\end{align}
where $\gamma$ can be adjusted to balance the original objective with the goal of a low-rank solution. The key idea is simple: if \eqref{SDP2} has a rank-one solution (thanks to the regularization), then one has found the global optimum of a QCQP that is not exactly equivalent to the original QCQP of \eqref{QCQP2}, but is similar to it. (The smaller $\gamma$ is, the more similar the two problems are.) Intuitively, the global optimum of a slightly modified problem is likely to be a high-quality solution of the original problem, an idea that has been supported in previous numerical experiments across optimization theory, ptychography, and optimal power flow~\cite{Goemans1995,Acikmese2007,Luo2010,Sojoudi2012,madani2014convex,Horstmeyer2015}, and is further supported by our results in the next section.

\Eqrefrange{QCQP2}{SDP2} are key results of this paper. \Eqref{QCQP2} is the culmination of an exact transformation of a wide range of physical design problems, with objectives subject to differential-equation constraints, into sparse QCQPs. This connection to a common mathematical structure leads to \eqrefs{SDP}{SDP2}, the first of which is a convex problem that is guaranteed to identify bounds for the problem of interest, and the second of which offers a new approach to physical design, with the possibility for circumventing low-quality local optima. In the next two sections, we perform numerical experiments that demonstrate the power of these transformations.

\section{Large-scale computational bounds}
In this section, we demonstrate the utility of our sparse-QCQP-based SDP, of \eqref{SDP}, for enabling computational bounds of large-scale systems for which such analysis is impossible with the current state-of-the-art methods. ``Metasurfaces,'' in which one patterns a wavelength-scale-thickness material to achieve high-performance optics functionality in compact form factors, offer a compelling example~\cite{Yu2014,Lalanne2017,aieta2015multiwavelength,Arbabi2015,avayu2017composite,Kamali2018,Chung2020}. A metalens is a metasurface that focuses light to a single focal spot, and one typically wants a maximally efficient metalens with a diameter significantly larger than the free-space optical wavelength. In \figref{metalens}, we consider a two-dimensional metasurface with diameter $d$, thickness $t$, refractive index $n=\sqrt{2}$, and numerical aperture $\textrm{NA} = 0.9$, and pose a fundamental question: what is the maximum possible efficiency any designable pattern could achieve? To answer this question, we formulate the Maxwell-constrained design problem of \eqref{opt_prob1}, transform the problem to the sparse QCQP of \eqref{QCQP2}, and relax the problem to the SDP of \eqref{SDP}. We exploit the sparsity of the differential operators using the clique-decomposition technique outlined in \figref{sparsity}, with the open-source software package \texttt{SparseCoLO}~\cite{Kim2011}, to dramatically reduce the size of the SDP and enable computation of large-scale bounds. \figref{metalens}(b,c) show bounds on the field intensities as a function of metalens diameter. One can see that the maximum field intensity depends sensitively on the metasurface thickness, and is well below the ``unitary bound'' arising from imposing unitarity constraints on the scattering matrix (cf. SM). A natural question is whether these bounds are achievable, and we include in \figref{metalens}(b) four notable data points (black markers): designs taken from the first singular vectors of the SDP solutions can achieve real performance levels almost exactly coinciding with the bounds themselves; the designs found by this SDP procedure are shown in the blue inset patterns. Crucially, the sparsity enables computation at very large scale sizes. Whereas integral-equation formulations of QCQP bounds, as in Refs.~\cite{Kuang2020,Molesky2020}, require computational times that scale with the fourth power of the metasurface diameter, the differential-equation-based bounds scale \emph{linearly}, enabling bounds for devices with diameters up to 100 free-space wavelengths in size.

\section{A new approach to design}
\begin{figure*}[tb]
    \centering
    \includegraphics[width=.85\textwidth]{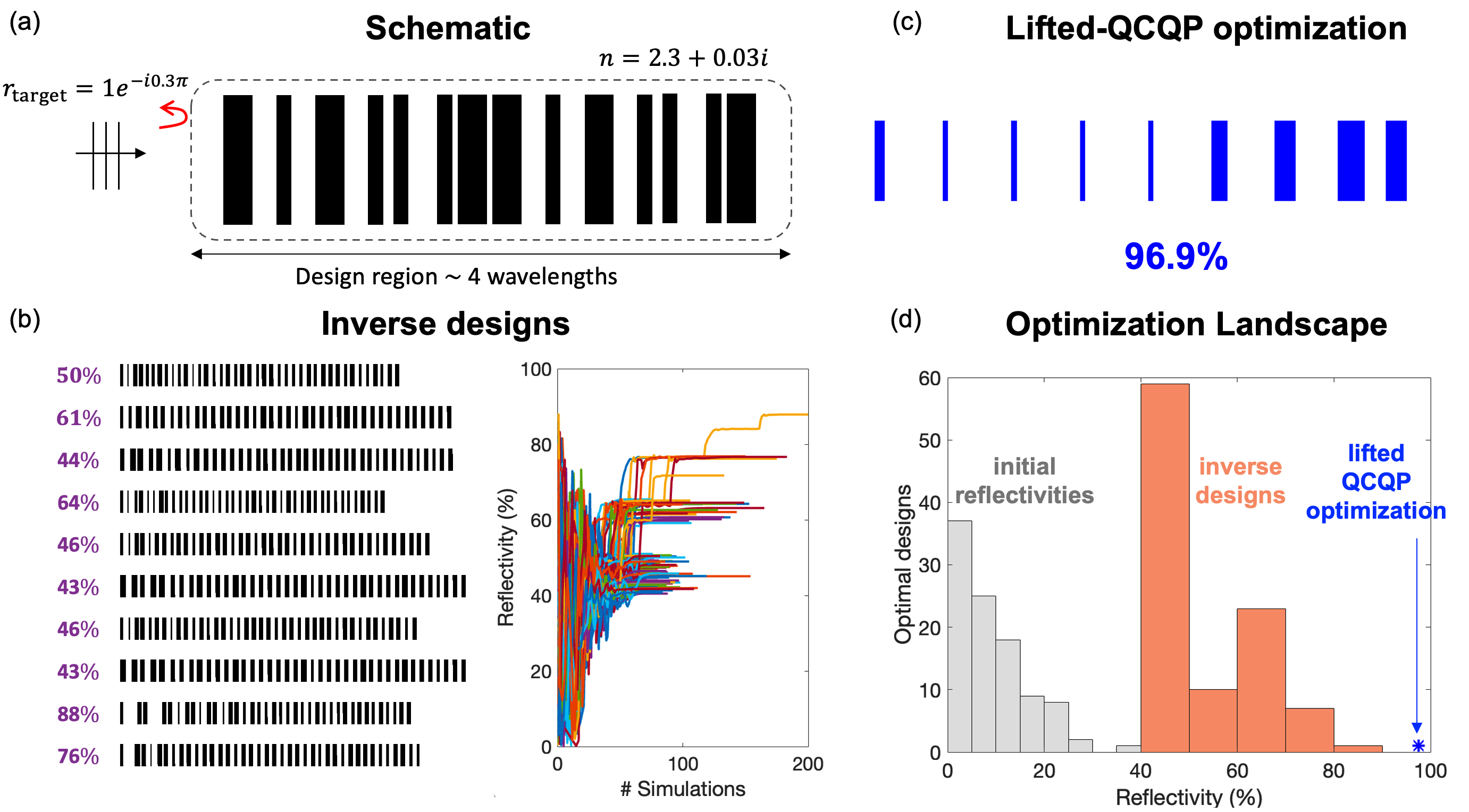}
            \caption{(a) Schematic of a simple design problem with significant wave-interference effects: achieving large reflection, with a specific phase, from a multilayer stack of a slightly lossy material and an air background. (b) Inverse designs using interior-point methods to use adjoint-based gradient information can easily be rapped in local optima; the bulk of the optimization runs converge to 40--50\% locally optimal efficiencies, while the best run, out of 100, finds a design with 88\% reflectivity. (c,d) A single, iterative optimization using the augmented-SDP of \eqref{SDP2} leads to a design (design widths in SM) that achieves 96.9\% efficiency, far outperforming even the best of the 100 inverse-design runs.}
    \label{fig:design}
\end{figure*}
Having shown the success of the SDP approach to large-scale bounds, we now turn to the question of whether our lifted-QCQP framework can lead to a new design approach altogether. The idea is to use the regularized form of the lifted QCQP, in \eqref{SDP2}, to promote a rank-one solution of the lifted problem, which will be a global solution of a closely related QCQP. To test this approach, we consider a canonical example that showcases the difficulty of physical design in the presence of wave-interference effects: a multilayer nanophotonic film, with waves scattering back and forth between the many possible layers. We consider a maximum-reflection problem as shown in \figref{design}, with a design region of multiple (in this case, four) free-space wavelengths, to achieve perfect reflection (reflection magnitude 1) with a specific reflection phase (arbitrarily chosen to be $-0.3\pi$). We consider a slightly lossy material (with refractive index $n=2.3+0.03i$, with a loss tangent typical of lossy materials in the visible); otherwise, any random collection of enough scatterers could generate perfect reflection (as in white paint). First, we run 100 full optimizations of ``inverse design''~\cite{Jensen2011,Miller2012}; we use a typical implementation of inverse design, with the widths of the films as the degrees of freedom, and adjoint-based simulations to speed up the gradient calculations. We use a standard quasi-Newton ``interior-point'' algorithm of \texttt{Matlab}, with default convergence criteria, and we start with about half of the region as air and half as the material, with 80 layers whose widths are randomly chosen between 0 and \SI{100}{nm}. (With significantly fewer than 80 layers, the optimizations tend to fall into even more low-quality local optima, while going well above 80 imposes a significant computational burden. The choice of 80 optimizes the tradeoffs between these competing effects.) Results from the 100 optimization runs are shown in \figref{design}(b), with the bulk of the optimizations converging to designs with reflectivity between 40--50\%, and the very best single design achieving only 88\% reflectivity. Ten designs---eight random, alongside the two best---are shown on the left-hand side of \figref{design}(b), showing the many different locally optimal designs that are discovered. Clearly one can see that gradient-based algorithms are prone to converging to local optima, especially for a problem with significant wave reflections.

To tackle this problem with an augmented SDP, per \eqref{SDP2}, we adapted an algorithm developed in \citeasnoun{Liu2019} for QCQPs that arise in optimal power flow applications. The rank of a matrix can be written as a summation of step functions applied to the singular values of the matrix; in \citeasnoun{Liu2019}, they use a smooth approximation for the step function to construct the rank-penalization function $\mathcal{R}(\XX)$. This function is well-suited for a majorization--minimization (MM) algorithm that is tailored to concave objectives subject to a convex constraint set. The MM algorithm applied as in \citeasnoun{Liu2019} iteratively solves semidefinite programs with rank-penalized objectives until a locally optimal rank-one solution is found, at which point a hyperparameter can be tweaked if one wants a certificate of global optimality. We find the latter step unnecessary, with a single (iterative) local optimum repeatedly showing quite good performance. Running the algorithm on the lossy-material high-reflection optimization discussed above leads to the result is shown in \figref{design}(c): a design that achieves 96.9\% reflectivity (and the correct phase), in a single optimization run, without any fine-grained hyperparameter tuning. One can see the difference between this approach and the inverse design approach in \figref{design}(d), where the 100 initial reflectivities and locally optimal inverse-design reflectivities are shown in gray and orange, respectively, and the single lifted-QCQP-based optimization is shown in the blue marker. Clearly, the lifted-QCQP approach leads to a better design, with a more robust design process. One cannot directly compare the number of simulations used in each approach, as the lifted-QCQP optimization does not do any ``simulations'' at each iteration, but the typical inverse-design optimization required 21 seconds on a 2021 Macbook Pro, for a total of 35 minutes for the 100 optimization runs, with a maximum reflectivity of 88\%. The lifted-QCQP optimization took 5 minutes on the same computer, leading to a design with 96.9\% reflectivity. (The reported time is the full runtime, with all subroutines and iterations included.) Moreover, the design discovered in the lifted-QCQP approach appears to be a quite plausible global optimum for this problem. Having seen the design, one can explain the physics of its operation: the spacings between consecutive films are nearly exactly half a wavelength, to create constructive interference in the reflected wave. Moreover, the films show a thickness gradation, which is sensible given the lossiness of the material: the first layers are likely to interact with many scattered waves, and should be thin to avoid material losses, while the rear layers simply need to reflect the waves at any cost and can be thicker as a result. (The thicker first layer likely sets the correct phase.) Hence this result shows not only very good performance, but is suggestive of discovering a near-globally-optimal design, as anticipated in the theoretical derivations of \secref{convex}.

\section{Discussion and conclusion}
In this paper, we have connected design problems subject to the dynamics of the linear differential equations of physics to modern convex optimization theory. By rewriting these design problems as conservation laws defined by differential operators, we identify a unique structure pervasive across physical design problems: they can be transformed to sparse QCQPs. Sparse QCQPs can be lifted to higher dimensions, where their SDP relaxations offer a general framework for large-scale computational bounds. Tantalizingly, regularizing the SDP can lead to efficient techniques for identify designs that may often approach the global optima, as we demonstrated for an electromagnetic design problem subject to significant amounts of speckle and wave-interference effects.

Many areas of science and engineering, including optimal power flow~\cite{madani2014convex,Sojoudi2012}, ptychography~\cite{Horstmeyer2015,Yurtsever2021}, VLSI design and Ising problems~\cite{Barahona1988}, and more, have benefited from connections to semidefinite programming going back two decades~\cite{Luo2010}. It had seemed that wave-physics and related physical design problems simply had unfavorable mathematical structure, where such techniques would not work, but our approach shows that even wave-based design problems can fit into this paradigm. One concern might be the significant computational cost of standard SDP solvers---they have polynomial runtimes, but the polynomial is typically 3.5 or 4 (i.e., runtime scales as $N^{3.5}$ or $N^4$ for $N$ degrees of freedom), which prevents the application of SDPs to large-scale optimization problems~\cite{Luo2010}. Auspiciously, there has been significant recent progress in computational algorithms for semidefinite programming~\cite{Burer2003,Arora2016,Boumal2016,Ding2021,Yurtsever2021}; in one recent work~\cite{Yurtsever2021}, SDPs with $10^7$ designable degrees of freedom, corresponding to $10^{14}$ matrix variables, were solved. Moreover, whereas we used the \emph{chordal sparsity} in tandem with clique decompositions that are possible when one dimension of the problem is significantly larger than any others, newer SDP techniques such as those of \citeasnoun{Yurtsever2021} can exploit broader sparsity characteristics, beyond just chordal sparsity. These improvements may be readily adaptable to our physical-design SDPs.

Looking forward, there is ample opportunity to apply this technique broadly across many applications in photonics, quantum control, elasticity, and more. In addition to the scalability improvements enabled by the differential-equation formulation, discussed above, another feature of our approach is that it can work seamlessly with the many open-source and commercial differential-equation solvers in widespread use. As an example, in the {\SM}, we show that one can create an SDP directly from the governing matrix equations created by \texttt{Comsol}~\cite{Comsol} and successfully identify the correct bounds. This suggests the possibility for our approach to become widely adopted across common differential-equation solvers.

An interesting contrast between our approach versus ``inverse design'' and deep-learning approaches is that the latter are ``structure-blind:'' aside from changes in the oracle functions to be called, the algorithms are not meaningfully modified for different design problems (linear, nonlinear, chaotic, etc.). Our approach specifically applies to physical design problems with bilinear structure in the field and geometric degrees of freedom (which are quite widespread), and cannot be generically applied to any problem. This is consistent with the ``no free lunch'' theorem~\cite{Wolpert1997}: there cannot be a single optimization method that is superior to others on all optimization problems. Our results suggest that a bespoke approach, with algorithms tailored to the underlying mathematical structure of the design problem of interest, may lead to a suite of design approaches that are superior to the structure-blind algorithms used today.

\section{Acknowledgments}
We thank Javad Lavaei and Somayeh Sojoudi for helpful discussions regarding sparse-QCQP formulations of optimal power flow. This work was supported by the Army Research Office under Grant No. W911NF-19-1-0279 and by the Air Force Office of Scientific Research under Grant No. FA9550-22-1-0393.

\bibliography{diff_bounds}

\end{document}

\title{Supplementary Materials:\\Many Physical Design Problems are Sparse QCQPs}

\author{Shai Gertler}
\affiliation{Department of Applied Physics and Energy Sciences Institute, Yale University, New Haven, CT 06520, USA}
\author{Zeyu Kuang} 
\affiliation{Department of Applied Physics and Energy Sciences Institute, Yale University, New Haven, CT 06520, USA}
\author{Colin Christie}
\affiliation{Department of Applied Physics and Energy Sciences Institute, Yale University, New Haven, CT 06520, USA}
\author{Owen D. Miller}
\affiliation{Department of Applied Physics and Energy Sciences Institute, Yale University, New Haven, CT 06520, USA}

\begin{abstract}

\end{abstract}

\maketitle
\tableofcontents

\newpage

\section{Detailed constraint-formulation methodology}
In this section, we provide further details on the form of the quadratic constraints introduced in the main text. We start with the differential equation
\begin{align}
    \LL(\chi) \psi - \xi = 0.
\end{align}
We will assume a discretization for which there are $N$ designable points in space/time, there are $M$ non-designable points in space/time (e.g. PML regions, background materials, etc.), and that there are $P$ vector (polarization, Hilbert-space, etc.) degrees of freedom at each point in space/time. Then $\psi$ and $\xi$ both have sizes $(N+M)P\times 1$, and $\LL$ is a matrix of size $(N+M)P \times (N+M)P$. We will take the typical case in which the control field $\chi$ can be controlled at each designable space/time point, but does not have independent levers of control along the polarization/Hilbert/etc. axis. (For example, we are assuming one cannot independently control the material properties of each polarization at a given point in space in a nanophotonics problem. Such control in fact slightly reduces the size of the ultimate QCQPs, and slightly simplifies the analysis below.)

First, let us tackle what happens at a designable point in the domain. If we denote this point $i$, then 
\begin{align}
    \left(\LL(\chi_{1,2}) \psi - \xi\right)\big\rvert_i
\end{align}
is a $P\times 1$ vector for each of $\chi_1$ and $\chi_2$. We want to enforce the condition that
\begin{align}
    \textrm{either } \left(\LL(\chi_{1}) \psi - \xi\right)\big\rvert_i = 0 \textrm{  or  } \left(\LL(\chi_{2}) \psi - \xi\right)\big\rvert_i = 0,
\end{align}
where the zeros are the right-hand sides are the zero vectors. Consider two vectors $a$ and $b$, both of size $P\times 1$, for which we want to enforce the condition that either one of the vectors is the zero vector. How can we do that? The key is to enforce
\begin{align}
    a_j^* b_k = 0, \qquad \forall j,k,
    \label{eq:quadab}
\end{align}
i.e., to force each pairwise inner product to be zero. This expression says that for each $j,k$, one of the two entries must be zero. (Or both, but that is typically forbidden by the form of $\LL(\chi)$.) The only way for this condition to be met for all $j$ and $k$ is for all of the entries of one of the vectors to be zero. This can be proven by contradiction. Suppose both vectors are nonzero---that is, they each have at least one element that is not zero---and that they satisfy \eqref{quadab}. Then choose $j$ and $k$ to be the indices of nonzero elements in each vector. Their product is not zero, contradicting the requirement that \eqref{quadab} be satisfied. Thus we see that \eqref{quadab} requires (at least) one of the vectors to be zero, which is exactly the either--or condition that we desire.

We can alternatively write the condition of \eqref{quadab} in matrix form,
\begin{align}
    a^\dagger \DD_{jk} b = 0, \qquad \forall j,k,
\end{align}
where $\DD_{jk}$ is a $P\times P$ matrix with all zeros except a single entry of 1 in its $j,k$ element. Now let us return to the differential equations of above. We can just directly substitute $\left(\LL(\chi_{1}) \psi - \xi\right)\big\rvert_i$ for $a$, and similarly but with $\chi_2$ for $b$, to get
\begin{align}
    \left(\LL(\chi_{1}) \psi - \xi\right)^\dagger \big\rvert_i \DD_{jk} \left(\LL(\chi_{2}) \psi - \xi\right)\big\rvert_i = 0.
    \label{eq:quadDiff}
\end{align}
Finally, we clean up the notation a bit. In \eqref{quadDiff}, $i$ is in index for the space/time point, while $j$ and $k$ are indices in the polarization/Hilbert space. We can wrap all of these into a single index $i$ that iterates over all space/time points and all polarization/Hilbert-space combinations; we can also enlarge $\DD$ to be a fully $(N+M)P \times (N+M)P$ matrix (instead of smaller $P\times P$ matrices iterated $N$ times), now just with all entries of the larger matrix being zero except at single $1$ at the $i^{\rm th}$ entry. Then we can drop the evaluation of the vectors at $i$ (since multiplication by the entire $\DD$ matrix effectively does this already), and we can write \eqref{quadDiff} in exactly the same form as expressed in the main text,
\begin{align}
    \left[\LL(\chi_{1}) \psi - \xi\right]^\dagger \DD_i \left[\LL(\chi_{2}) \psi - \xi\right] = 0.
    \label{eq:quadEq}
\end{align}
Next, we tackle the \emph{non-designable} points in the design space. At each of those points, there are not two possibilities for $\chi$, but instead only a single one---whatever the background material is at that point in the domain. Hence, we enforce $MP$ linear constraints of the form 
\begin{align}
    \left[\LL(\chi_{i}) \psi - \xi\right]\big\rvert_i = 0.
    \label{eq:linND}
\end{align}
However, once the problem is lifted and relaxed in a higher-dimensional space, it is not clear that these constraints will be binding, and indeed it has been found quite useful for some QCQPs to add seemingly redundant constraints to their formulation; those constraints are not redundant in the lifted, relaxed space~\cite{Poljak1995}. We find it helpful to add $MP^2$ quadratic forms exactly equivalent to \eqref{quadEq} at each nondesignable point, except that now \emph{both} terms have $\chi_i$ in them:
\begin{align}
    \left(\LL(\chi_{i}) \psi - \xi\right)^\dagger \DD_i \left(\LL(\chi_{i}) \psi - \xi\right) = 0.
    \label{eq:quadND}
\end{align}
The constraints of \eqrefrange{quadEq}{quadND} form all of the constraints of the system. In total, there are $(N+M)P^2$ quadratic constraints, and $MP$ linear constraints.

What if one has multilevel level controls? If there were three possible control-field values, $\chi_1$, $\chi_2$, and $\chi_3$, then the natural analog of the approach above is to multiply each of the three expressions, $\LL(chi_i)\psi - \xi$, together, to form a single constraint, which would then be a third-order polynomial in $\psi$. More generally, for $N$ possible control-field values, one would form $N^{\rm th}$-order polynomial constraints by this approach. It is well-known that optimization problems with polynomial objectives and polynomial constraints are a subclass of QCQPs---cf. Sec.~2.1 of \citeasnoun{Park2017}. The key idea is the extra variables can be introduced to reduce the order of the problem, at the expense of additional constraints. For example, if a constraint is $y^3 = 0$, one equivalently enforce $xy = 0$ and $x = y^2$; in this example, two second-order (at most) constraints are equivalent to the single third-order constraint.

\section{Interpreting the quadratic constraints as conservation laws}
In this section, we justify our statement that a constraint of the form 
\begin{align}
    \left[\LL(\chi_{1}) \psi - \xi\right]^\dagger \DD_i \left[\LL(\chi_{2}) \psi - \xi\right] = 0
    \label{eq:quadEq2}
\end{align}
represents a conservation law. To do so, we first go back to the continuous representation, in which $\LL$ is an operator and $\psi$ is a vector field, and rewrite \eqref{quadEq2} as an overlap integral. The matrix $\DD_i$ in \eqref{quadEq2} has a single nonzero entry; its purpose is to isolate a single space/time point and specific polarizations (Hilbert-space indices) for the two equations. A similar isolation of specific space/time/polarization/etc. locations can be achieved in continuous space by taking an inner product with a function $v_i$ that encodes the desired localization. Hence we can write
\begin{align}
    \left\langle \LL(\chi_{1}) \psi - \xi, \LL(\chi_{2}) \psi - \xi\right\rangle_{v_i} = 0,
    \label{eq:quadEq3}
\end{align}
where $\langle x,y \rangle_v = \int x^\dagger v y$, and $\dagger$ denotes the adjoint operator. To simplify notation below, we will assume $\chi_2$ is always zero, and rewrite $\chi_1$ as $\chi_c$, as the ``background'' control field can always be shifted to make this true, such that the quadratic constraint is:
\begin{align}
    \left\langle \LL(\chi_c) \psi - \xi, \LL(0) \psi - \xi\right\rangle_{v_i} = 0.
    \label{eq:quadEq4}
\end{align}
In the following subsections, we use the explicit forms of $\LL(\chi)$ in electromagnetism and quantum control to show that \eqref{quadEq4} can be interpreted as a conservation law.

\subsection{Electromagnetic design}

We consider Maxwell's equation describing a monochromatic field at frequency $\omega$ in a nonmagnetic material (i.e., the relative permeability is $\mu_r=1$ everywhere in space).
The electric field $\textbf{E}(x)$ can be written as
\begin{align}
    \Big[\nabla \times \nabla \times - \left(1+\chi(\textbf{x})\right)\omega^2 \Big]\textbf{E}(\textbf{x})  = i\omega \textbf{J}(\textbf{x}).
    \label{eq:EM}
\end{align}
where $\textbf{J}(\textbf{\textbf{x}})$ are source currents, $\chi$ is the (designable) susceptibility, and we use dimensionless units where the vacuum permittivity and permeability are both 1. In line with the discussion above we take $\chi$ to be allowed two values, 0 or $\chi_c$. Then,
\begin{align}
    \LL(\chi_c) = \nabla \times \nabla \times - \left(1+\chi_c\right) \omega^2.
\end{align}
To simplify our notation, we will define a background-$\LL$ operator,
\begin{align}
\LL_0 = \nabla \times \nabla \times - \omega^2,
\end{align}
so that $\LL(\chi_c) = \LL_0 - \chi_c\omega^2$. Then \eqref{quadEq4} can be written:
\begin{align}
    \braket{ \LL_0 \textbf{E} - i\omega\textbf{J} ,  \LL_0 \textbf{E} -\omega^2\chi_c \textbf{E} -i\omega\textbf{J}}_{v_i}=0.
\end{align}
Multiplying by a factor ${\chi_c^{-1}}/{\omega^2}$ and rearranging leaves us with the constraint:
\begin{align}
    \braket{\left( \LL_0 \textbf{E} - i\omega\textbf{J}\right) , \frac{\chi_c^{-1}}{\omega^2}\left( \LL_0 \textbf{E} -i\omega\textbf{J}\right)}_{v_i} - \braket{ \LL_0 \textbf{E} , \textbf{E}}_{v_i} -i\omega \braket{\textbf{J} , \textbf{E}}_{v_i} =0,\quad {i=1,2,3\dots}
    \label{eq:constEM}
\end{align}
To see the equivalence of this form to a power conservation law, we introduce the ``polarization density''
\begin{align}
    \textbf{P} = \chi\textbf{E} = \frac{1}{\omega^2} \left( \left[\nabla \times \nabla \times - \omega^2 \right] \textbf{E} - i\omega\textbf{J} \right).
\end{align}
Using this definition in the constraint we have derived in Eq. (\ref{eq:constEM}) we have
\begin{align}
    \omega^2 \braket{\textbf{P} , {\chi_c}^{-1}\textbf{P}}_{v_i} - \braket{\nabla\times \nabla\times \textbf{E},\textbf{E}}_{v_i} + \omega^2 \braket{ \textbf{E}, \textbf{E}}_{v_i} -i\omega \braket{\textbf{J} , \textbf{E}}_{v_i} =0,\quad {i=1,2,3\dots}
    \label{eq:const_max}
\end{align}
We can further simplify the second term; using Maxwell's equation ${\nabla \times \textbf{E}=i\omega\textbf{H}}$ and the vector identity \\ ${\textbf{A}\cdot(\nabla\times \textbf{B})=\textbf{B}\cdot(\nabla\times\textbf{A})-\nabla\cdot(\textbf{A}\times\textbf{B})}$, we have
\begin{align}
\begin{split}
    \braket{\nabla\times \nabla\times \textbf{E},\textbf{E}} &= -i\omega \braket{\nabla\times \textbf{H},\textbf{E}}\\ 
    &= -i\omega \braket{\textbf{H},\nabla\times \textbf{E}}+i\omega \int {\rm d}V \ \nabla\cdot\left(\textbf{E}\times\textbf{H}^*\right)\\ 
    &= \omega^2 \braket{\textbf{H}, \textbf{H}}+i\omega \oint {\rm d}A \ \left(\textbf{E}\times\textbf{H}^*\right) \cdot \mathbf{n},
\end{split}
\end{align}
where in the last step we used the divergence theorem to express  the second term as a surface integral.
Next, we turn to the first term in Eq. (\ref{eq:const_max}); at points where ${\chi(x)=\chi_c}$ we know that ${\textbf{E}=\chi_c^{-1}\textbf{P}}$.
This is not true in regions where ${\chi(x)=0}$, however, in these regions $\textbf{P}=0$. 
Hence, we can rewrite the inner product as ${\braket{\textbf{P} , {\chi_c}^{-1}\textbf{P}}=\braket{\textbf{P} , \textbf{E}}}$. 
All together, this leaves us with
\begin{align}
    i\omega \braket{\textbf{P} ,\textbf{E}}_{v_i} + i\omega \braket{ \textbf{E}, \textbf{E}}_{v_i} - i\omega \braket{\textbf{H},\textbf{H}}_{v_i} +\oint {\rm d}A \left(\mathbf{E}\times\mathbf{H}^*\right)\cdot \mathbf{ n} + \braket{\textbf{J} , \textbf{E}}_{v_i} =0,\quad {i=1,2,3\dots}
\end{align}
which is precisely the complex Poynting theorem for the case of non-magnetic material \cite{jackson1999classical}.

\subsubsection*{Six-vector notation}

We can write a general six-vector form for Maxwell's equations; denoting the fields $\psi(x) = \begin{pmatrix} \textbf{E} & \textbf{H} \end{pmatrix}^T$ and the current sources $\xi(x) = \begin{pmatrix} \textbf{J} & \textbf{K} \end{pmatrix}^T$, we have
\begin{equation}
    \begin{pmatrix} 0 & \nabla\times \\ -\nabla\times & 0\end{pmatrix} \psi + i\omega \psi + i\omega\chi \psi = \xi,
\label{eq:Max}
\end{equation}
where $\chi$ is the material susceptibility tensor, which can have both electric and magnetic (and magnetoelectric) components.
We assume the susceptibility tensor can take one of two values at each point in space ${\chi(x) \in\{0,\chi_c\}}$, and we can define the operator
\begin{align}
     \LL_0 = \begin{pmatrix} 0 & \nabla\times \\ -\nabla\times & 0\end{pmatrix} +i\omega\mathbb{I},
\end{align}
such that we have the constraint
\begin{align}
    \braket{\left( \LL_0\psi - \xi\right) , \frac{\chi_c^{-1}}{i\omega}\left( \LL_0\psi - \xi\right)}_{v_i} + \braket{ \LL_0\psi , \psi}_{v_i} - \braket{\xi , \psi}_{v_i} =0,\quad {i=1,2,3\dots}
    \label{eq:constEM6vect}
\end{align}
Defining a ``polarization field''
\begin{align}
\phi= \chi \psi = \frac{i}{\omega}\left( \LL_0\psi - \xi\right),
\end{align}
we have
\begin{align}
    -i\omega\braket{\phi , \chi_c^{-1}\phi}_{v_i} + \braket{\begin{pmatrix} 0 & \nabla\times \\ -\nabla\times & 0\end{pmatrix} \psi, \psi}_{v_i} -i \omega\braket{\psi , \psi}_{v_i} - \braket{\xi , \psi}_{v_i}=0,\quad {i=1,2,3\dots}
\end{align}
As we saw earlier, in regions where $\chi=\chi_c$ we have $\chi_c^{-1}\phi=\psi$, and where $\chi=0$ the polarization vanishes ($\phi=0$), such that we can rewrite the first term
\begin{align}
    -i\omega\braket{\phi,\psi}_{v_i} + \braket{\begin{pmatrix} 0 & \nabla\times \\ -\nabla\times & 0\end{pmatrix} \psi, \psi}_{v_i} -i \omega\braket{\psi , \psi}_{v_i} - \braket{\xi , \psi}_{v_i}=0,\quad {i=1,2,3\dots}
    \label{eq:6vect_Poynt}
\end{align}
Taking the real part of Eq. (\ref{eq:6vect_Poynt}), we have a statement of the real Poynting's theorem \cite{chew2008integral}
\begin{align}
    \text{Im}\left[\omega\braket{\phi , \psi}_{v_i}\right] + \underbrace{\text{Re}\left[\braket{\begin{pmatrix} 0 & \nabla\times \\ -\nabla\times & 0\end{pmatrix} \psi, \psi}_{v_i}\right]}_{= \text{Re} \left[\oint {\rm d}A \left(\mathbf{E}\times\mathbf{H}^*\right)\cdot \mathbf{ n}\right]} - \underbrace{\text{Re}\left[i \omega\braket{\psi , \psi}_{v_i}\right]}_{=0} - \text{Re}\left[\braket{\xi , \psi}_{v_i}\right]=0,\quad {i=1,2,3\dots}
\end{align}
We can follow a similar procedure to obtain the imaginary part of Poynting's theorem. Interestingly, it does \emph{not} arise simply from taking the imaginary part of \eqref{6vect_Poynt}.

\subsection{Quantum control}
In this subsection we show that our quadratic-constraint formulation of quantum control problems also represents conservation laws. We start with a Hamiltonian of the form
\begin{align}
    H(t) = H_0(t) + \chi(t) H_c(t),
    \label{eq:Htot}
\end{align}
where $H_0$ is the non-controllable part of the Hamiltonian, $H_c(t)$ is the controllable part, and $\chi(t)$ is the control trajectory to be designed. The Schrodinger equation describes the temporal dynamics of the time-evolution operator $ U(t,t_0)$ (for some initial time $t_0$), given by
\begin{align}
    i \hbar\frac{d}{dt} U(t,t_0) =  H(t) U(t,t_0).
    \label{eq:Schr}
\end{align}
In this case, rather than source term, we have an initial condition at time $t=t_0$.
Using Eqs. (\ref{eq:Htot}) and (\ref{eq:Schr}), we have 
\begin{align}
    \left[i\frac{d}{dt} -  H_0(t) \right] U(t,t_0) - \chi(t)  H_c(t) U(t,t_0) = 0.
\end{align}
where we have set $\hbar=1$. We assume the control parameter can take one of two values ${\chi(t) \in\{0,\chi_c\}}$, and denoting the operator
\begin{align}
     \LL_0 = i\frac{d}{dt} -  H_0,
\end{align}
we have the constraint:
\begin{align}
    \braket{ \LL_0 U(t,t_0) , \LL_0 U(t,t_0) - \chi_c H_c U(t,t_0) } =0.
\end{align}
Multiplying by ${H_c^{-1}}/{\chi_c}$ and rearranging leaves us with the constraint:
\begin{align}
    \braket{ \LL_0 U(t,t_0) , \frac{ H_c^{-1}}{\chi_c} \LL_0 U(t,t_0)} - \braket{ \LL_0 U(t,t_0) , U(t,t_0) } =0.
    \label{eq:constQM}
\end{align}
Inserting the expression for $\LL_0$ into the term on the right-hand side and rearranging gives
\begin{align}
    \braket{ \frac{d}{dt} U(t,t_0) , U(t,t_0) } = i\braket{ \LL_0 U(t,t_0) , \frac{ H_c^{-1}}{\chi_c} \LL_0 U(t,t_0)} + i\braket{U(t,t_0) ,H_0^\dagger U(t,t_0) }.
    \label{eq:QCcons}
\end{align}
Taking the real part of this equation leaves us with:
\begin{align}
    &\frac{d}{dt} \braket{ U(t,t_0) , U(t,t_0)} =\\
    &-2 \braket{ \LL_0 U(t,t_0) , \text{Im}\left[\frac{ H_c^{-1}}{\chi_c} \right] \LL_0 U(t,t_0) } + 2 \braket{ U(t,t_0) , \text{Im}\left[ H_0 \right] U(t,t_0)},
\end{align}
where ``$\Im$'' denotes the anti-Hermitian part of the operator, and we have used the fact that $\text{Im}\left[ H_0^\dagger \right] = -\text{Im}\left[ H_0 \right]$. In the case of real-valued control parameter and Hermitian operators, i.e., $\text{Im}\left[{ H_c^{-1}/\chi_c} \right] = \text{Im}\left[ H_0 \right] =0$, we have 
\begin{align}
    \frac{d}{dt} \braket{ U(t,t_0) , U(t,t_0)}=0,
\end{align}
which represents conservation of probability in a Hermitian (closed) system. In a non-Hermitian system, the term on the right-hand side would simply give the probability loss due to the non-Hermiticity. Taking the imaginary part of \eqref{QCcons} leads to a ``reactive'' probability conservation, in analogy with reactive power conservation in electromagnetism.

\section{Forming the SDP by lifting the QCQP}
In this section we provide the explicit transformations from the QCQP form of the problem to the SDP relaxation. This procedure is well-known, cf. \citeasnoun{Luo2010}, but we also include the formulation here, for reproducibility. We start by writing the QCQP:
\begin{equation}
    \begin{aligned}
    & \underset{\psi\in\CC^n}{\text{maximize}}
    & & \psi^\dagger \mathbb{A} \psi +  \Re\left(\beta^\dagger \psi\right) \\
    & \text{subject to}
    & &  \left[\LL(\chi_1) \psi - \xi\right]^\dagger \DD_i\left[\LL(\chi_2) \psi - \xi\right] = 0,\myforall i \in I_d, \\
    & & & \left[\LL(\chi_{i}) \psi - \xi\right]\big\rvert_i = 0 ,\myforall i \in I_{\rm PML}, \\
    & & & \left(\LL(\chi_{i}) \psi - \xi\right)^\dagger \DD_i \left(\LL(\chi_{i}) \psi - \xi\right) = 0,\myforall i \in I_{\rm PML},
    \end{aligned}
    \label{eq:SM-QCQP}
\end{equation}
where $I_d$ is the set of indices in the designable domain, $I_{\rm PML}$ is the set of indices in the PML, and $\CC^n$ refers to the set of $n$-dimensional complex-valued vectors. Before any transformations, we can compactify our notation:
\begin{equation}
    \begin{aligned}
    & \underset{\psi\in\CC^n}{\text{maximize}}
    & & \psi^\dagger \mathbb{A} \psi +  \Re\left(\beta^\dagger \psi\right) \\
    & \text{subject to}
    & &  \psi^\dagger \BB_i \psi + \Re\left(v_i^\dagger \psi\right)  = \gamma_i,\myforall i \in I_{\rm quad}, \\
    & & & \Re \left[ u_i^\dagger \left( \LL(\chi_{i}) \psi \right) \right] = \delta_i ,\myforall i \in I_{\rm lin}.
    \end{aligned}
    \label{eq:SM-QCQP2}
\end{equation}
To arrive at \eqref{SM-QCQP2}, we use four steps. First, we combine the quadratic constraints of \eqref{SM-QCQP} into a single form, differentiated only by index $i$, in the now larger set $I_{\rm quad}$. Second, we form the linear constraints by multiplying over a basis set of vectors $u_i$. Third, we define $\delta_i = \Re\left(u_i^\dagger \xi\right)$. Finally, we take the real parts of the quadratic and linear constraint equations of \eqref{SM-QCQP}, and we account for the imaginary parts by allowing for $\DD_i$ and $u_i$ to take unit entries or their complex counterparts (i.e., multiplied by the imaginary unit).

The first step to simplifying \eqref{SM-QCQP2} is to \emph{homogenize} the problem, which means to transform the linear terms in the QCQP to quadratic terms, so that all terms in \eqref{SM-QCQP} are either quadratic or scalar. We can do this by introducing a new slack variable $s$ into the objective and constraints, multiplying the linear terms. This slack variable must have absolute value 1, but it can be positive or negative; a negative one value would be compensated in $\psi$ by multiplying by -1. With this new slack variable, the problem can be written
\begin{equation}
    \begin{aligned}
    & \underset{\psi\in\CC^n, |s|^2 = 1}{\text{maximize}}
    & & 
    \begin{pmatrix}
        \psi \\
        s
    \end{pmatrix}^\dagger 
    \begin{pmatrix}
        \AA & \frac{\beta}{2} \\
        \frac{\beta^\dagger}{2} & 0
    \end{pmatrix}
    \begin{pmatrix}
        \psi \\
        s
    \end{pmatrix} \\
    & \text{subject to}
    & & 
    \begin{pmatrix}
        \psi \\
        s
    \end{pmatrix}^\dagger
    \begin{pmatrix}
        \BB_i & \frac{v_i}{2} \\
        \frac{v_i^\dagger}{2} & 0
    \end{pmatrix}
    \begin{pmatrix}
        \psi \\
        s
    \end{pmatrix}
    = \gamma_i,\myforall i \in I_{\rm quad}, \\
    & &  &
    \begin{pmatrix}
        \psi \\
        s
    \end{pmatrix}^\dagger
    \begin{pmatrix}
        0 & \frac{\LL^\dagger(\chi_i) u_i}{2} \\
        \frac{u_i^\dagger \LL(\chi_i)}{2} & 0
    \end{pmatrix}
    \begin{pmatrix}
        \psi \\
        s
    \end{pmatrix}
    = \delta_i ,\myforall i \in I_{\rm lin},
    \end{aligned}
    \label{eq:SM-QCQP3}
\end{equation}
in which all of the expressions are either quadratic in the combined variable $\begin{pmatrix}\psi & s\end{pmatrix}^T$, or scalar values. We can again make the notation more compact, and write this problem as a homogenized, complex-valued QCQP:
\begin{equation}
    \begin{aligned}
    & \underset{x}{\text{maximize}}
    & & x^\dagger \AA_0 x \\
    & \text{subject to}
    & & 
    x^\dagger \AA_i x = a_i \myforall i.
    \end{aligned}
    \label{eq:SM-QCQP4}
\end{equation}
To then convert this to a complex-valued SDP, we simply take the trace of each expression, which does not modify any values, then use the invariance of trace under cyclic permutations to form the product $x x^\dagger$, which we define to be a new positive semidefinite, rank-one matrix $X$. The equivalent problem for this matrix variable is
\begin{equation}
    \begin{aligned}
    & \underset{X}{\text{maximize}}
    & & \Tr\left(\AA_0 X\right) \\
    & \text{subject to}
    & & 
    \Tr(\AA_i X) = a_i \myforall i, \\
    & & & X \geq 0, \\
    & & & \operatorname{rank} X = 1.
    \end{aligned}
    \label{eq:SM-QCQP5}
\end{equation}
We can relax this problem by dropping the rank-one constraint, leaving the standard-form (complex-valued) SDP:
\begin{equation}
    \begin{aligned}
    & \underset{X}{\text{maximize}}
    & & \Tr\left(\AA_0 X\right) \\
    & \text{subject to}
    & & 
    \Tr(\AA_i X) = a_i \myforall i, \\
    & & & X \geq 0.
    \end{aligned}
    \label{eq:SM-SDP}
\end{equation}
In this form, the problem can be solved in any SDP solver. (If the solver does not accept complex-valued matrices, it is simple to double all dimensions and separate the real and imaginary parts of the vectors and matrices.)

\section{Graph theory for semidefinite programming}
\label{sec:quick_review}
Graphs distill real-world relations into abstract lines and vertices, for social networks in sociology, genetic networks in biology, communication networks in computer science, lattice structures in physics, and molecular chains in chemistry. Complex dynamics can simplify when occurring on graphs with special structure. In particular, chordal graphs represent a class of connections where many hard problems can be easily solved, including graph coloring, clique finding, and matrix factorization~\cite{Vandenberghe2015}. In this section, we review basics of graphs and chordal graphs, as well as theorems that foreground chordal graphs in sparse semidefinite programming. 
    
A graph comprises a set of vertices $V = \{v_1, v_2, ..., v_n\}$ and their connecting edges $E \subseteq V \times V$. The vertices and edges uniquely define the graph: $G(V,E)$. We consider undirectional graph whose edges are unordered pairs, denoted by curly brackets such as $\{v_i, v_j\}$.
If a series of edges leads one vertex back to itself, then these edges form a cycle.
Two vertices $v_i$ and $v_j$ are adjacent if there is an edge between them, i.e., if $\{v_i, v_j\} \in E$.
Vertices that are all adjacent to each other form a clique. 
A clique can expand upon admitting new vertices that are adjacent to all its existing members. If no such new vertices exist, the clique is then called a maximal clique of the graph.

Many large problems can be simplified if their graph is a chordal. A chordal graph is a graph in which every cycle of length four and greater has a chord (an edge between nonconsecutive vertices of the cycle). In other words, if you trace a cycle of four (or more) edges without finding a shortcut, then the graph is not a chordal graph. 
Any graph can be made chordal by adding extra edges. In the extreme case, supplying every possible edges guarantees a chordal graph, but then there can be no problem-size reduction. The art is to add as few edges as possible to make a graph chordal. This procedure is called chordal completion, often implemented via heuristic algorithms~\cite{amestoy1996approximate}.

Sparsity in matrices map to specific graph representations of the matrix data. Zeros in a symmetric matrix $\XX$ are missing edges in the corresponding undirected graph $G(V, E)$: $\{v_i, v_j\}\notin E \iff \XX_{ij} = \XX_{ji} = 0$. Two groups of symmetric matrices $\XX$ are of particular interest in sparse semidefinite programming. The first are matrices that are positive semidefinite and sparse, with sparsity pattern given by the graph $G(V, E)$:
\begin{equation}
    \SS_+^n(E,0) = \{\XX\in\SS^n_+\ |\ \XX_{ij}=\XX_{ji}=0\ \textrm{if}\ (i,j)\neq E\},
\end{equation}
where $\SS^n_+$ (without argument) refers to the set of positive definite symmetric matrices. The second group consists of matrices that are not necessarily positive semidefinite or sparse but can be ``completed'' into positive semidefinite matrices. To motivate this, consider an inner product $\Tr(\AA\XX)$ between the symmetric matrix $\XX$ and a sparse symmetric matrix $\AA$ whose sparsity is given by the graph $G(V, E)$.
The inner product multiplies matrices element-wise, so only the elements in $\XX$ that correspond to the edges in the graph $G(V, E)$ are involved in the computation of the inner product.  If the rest of the elements in $\XX$ can be altered to turn $\XX$ into a positive semidefinite matrix, say $\MM$, then we say $\XX$ belongs to a group of completable partial symmetric matrices:
\begin{equation}
    \SS_+^n(E,?) = \{\XX\in\SS^n\ |\ \exists \MM\geq 0, \MM_{ij} = \XX_{ij}  \ \forall (i,j)\in E\}
\end{equation}
The two matrix spaces, $\SS_+^n(E,0)$ and $\SS_+^n(E,?)$, are both convex cones and, in fact, are duals of each other~\cite{Vandenberghe2015}. Together, they constitute the basic matrix spaces in sparse semidefinite programming.

Crucially, if the graph $G(E, V)$ is a chordal graph, both types of matrices above can be decomposed into smaller matrices defined on the maximal cliques of the graph, $\{C_1, C_2, ..., C_l\}$.
Aiding this decomposition is a projection matrix $\TT_l$ that projects a matrix $\XX$ into its principle submatrix $\XX_l = \TT_l \XX \TT_l^\top$, and reads $(\TT_l)_{ij}=1$ if $C_l(i)=j$ and zero otherwise, where $C_l(i)$ is the $i$-th vertex in $C_l$.
If the matrix $\XX$ is positive semidefinite completable, i.e., $\XX\in\SS_+^n(E,?)$, then all the submatrices $\XX_l$ are all positive semidefinite, as dictated by Grone’s theorem~\cite{grone1984positive}:
\begin{equation}
    \XX \in \SS^n_+(E,?) \iff \XX_l = \TT_l \XX \TT_l^\top \in \SS^{|C_l|}_+.
    \label{eq:Grone}
\end{equation}
If the matrix $\XX$ is sparse positive semidefinite, i.e., $\XX\in\SS_+^n(E,0)$, then all the submatrices $\XX_l$ are positive semidefinite and uniquely expand the matrix $\XX$, as dictated by Agler’s theorem~\cite{agler1988positive}:  
\begin{equation}
    \XX\in\SS_+^n(E,0) \iff \XX = \sum_{l=1}^{p} \TT_l^\top \XX_l \TT_l, \ %\textrm{with\ } 
    \XX_l \in \SS^{|C_l|}_+.
    \label{eq:Agler}
\end{equation} 
Agler's theorem in Eq.~(\ref{eq:Agler}) allows one to decompose sparse semidefinite programs in their dual forms, where the matrix variable $\XX$ directly inherits the sparsity of the problem; Grone's theorem in Eq.~(\ref{eq:Grone}) allows one to decompose sparse semidefinite programs in their primal forms, where the the matrix variable $\XX$ does not inherit the sparsity but is multiplied by matrices that do. We discuss this second scenario in the next section.

\section{Fast semidefinite programming on chordal graphs}
\label{sec:fastSDP}

Semidefinite programs are challenging to solve for large-dimensional problems, unless there is sparsity to be leveraged. 
Sparsity mainly accelerates two operations in the semidefinite programming in \eqref{SM-SDP}: multiplying matrix $\XX$ by matrix $\AA_i$, and verifying the semidefinite constraint, $\XX\geq 0$, is satisfied. 
These two operations are the numerical bottleneck for most  semidefinite programming algorithms~\cite{Boyd2004, Luo2010}, both costing at least $O(n^3)$ per iteration for an $n$-dimensional problem.
To accelerate these two operations, one needs to consider the aggregate sparsity pattern of the semidefinite program in \eqref{SM-SDP}, which are the places where all the objective and constraint matrices, $\AA_i$, are zero. 
 The aggregate sparsity pattern is given by the sparsity pattern of the $\LL^\dagger \LL$ matrix in the QCQP in \eqref{SM-QCQP}, which is shown as the black and red squares in Fig.~1(b) in the main text for a number scenarios.
 Most sparsity can accelerate matrix-matrix multiplications by multiplying only the nonzero entries.
 On the other hand, only a special type of aggregated sparsity based on chordal graphs can accelerate the verification of the semidefinite constraint, which we explain below.

The major bottleneck of large-scale semidefinite programming is the verification of its semidefinite constraint, which is resolvable if the underlying sparsity pattern corresponds to a chordal graph~\cite{Vandenberghe2015}.
%Solving a semidefinite programming can be easy if its underlying sparsity pattern corresponds to chordal graphs. 
There is a unique mapping between sparsity pattern of symmetric matrices and undirectional graphs: an nonzero $(i,j)$ entry of the former maps to a edge between the $i$ and $j$ vertices of the latter. (See Section~\ref{sec:quick_review} for a quick review on graph theories.) 
A chordal graph, in particular, is a graph where all cycles of four or more vertices have a chord. 
As shown in the Fig.~1(a) of the main text, the sparsity pattern of the $\LL^\dagger\LL$ operator in our photonic problem does not necessarily correspond to a chordal graph but it can be made into one by adding extra chords (the green lines in Fig.~1(a)).
Chordal graph is desirable because it allows one to distribute certain complex operations into different ``cliques'', a set of vertices in the graph that are all connected to each other (as in Fig.~1(a)).
In particular, if the aggregate sparsity of the semidefinite program in \eqref{SM-SDP} can be completed into a chordal graph, then Grone's theorem in Eq.~(\ref{eq:Grone}) shows its semidefinite constraint $\XX \geq 0$ can be replaced by ``smaller'' semidefinite constraints on each of its cliques: $\XX[I_j, I_j] \geq 0$, where $I_j$ denoting the indices for the $j$th clique. The optimization problem of \eqref{SM-SDP} becomes
\begin{equation}
    \begin{aligned}
    & \underset{\XX\in\SS^n}{\text{maximize}}
    & & \Tr(\AA_0\XX) \\
    & \text{subject to}
    & &  \Tr(\AA_i\XX) = a_i, \quad \text{for } i = 1, 2, ..., m, \\
    & & & \ \XX[I_j, I_j] \geq 0, \quad \text{for } j = 1, 2, ..., p,
    \end{aligned}
    \label{eq:chap5-SDP-chordal}
\end{equation} 
where $p$ to denotes the total number of cliques in the chordal graph. 
The only difference between the decomposed semidefinite program and the original semidefinite program is the semidefinite constraints are now applied to smaller cliques.
As the semidefinite constraint is the main computational bottleneck that increases as $O(n^3)$ with the matrix size $n$, breaking a large $\XX$ into many smaller ones in \eqref{chap5-SDP-chordal} allows us to significantly accelerate the SDP computations.

\section{Implementation of the majorization-minimization algorithm}
To obtain a rank-1 solution to the QCQP, a proxy function $\mathcal{R}(\mathbb{X})$ for rank is incorporated into the objective as a penalty term to be minimized:
\begin{equation}
\label{opt_pen}
    \begin{aligned}
    & \underset{\mathbb{X}}{\text{maximize}}
    & & \Tr(\mathbb{A}\mathbb{X}) - \gamma\mathcal{R}(\mathbb{X}) \\
    & \text{subject to}
    & &  \Tr(\mathbb{B}_i\XX) = b_i, \quad \text{for } i = 1, 2, ..., m, \\
    & & & \ \XX \geq 0.
    \end{aligned}
\end{equation}
We successfully generated rank-1 solutions to Eq. (\ref{opt_pen}), following the methods introduced by Liu, Sun, and Tsang in \citeasnoun{Liu2019}, producing \emph{designs} from the modified-SDP formulation. Here we summarize the key steps of \citeasnoun{Liu2019} that we implemented.

The first step is to find an approximate representation of the rank. One way to write the rank is to count the number of nonzero singular values, which is given by
\begin{equation}
    \operatorname{rank}(\mathbb{X}) = \sum_i H(\sigma_i(\mathbb{X}))
\end{equation}
 where $\sigma_i(\mathbb{X})$ is the $i^{th}$ singular value of $\mathbb{X}$ and $H(x)$ is the Heaviside step function, equal to 1 if $x > 0$ and 0 if $x = 0$ for $x \geq 0$. $H(x)$ is non-differentiable and therefore cannot be used in gradient-based optimization, but it can be approximated by the smooth, differentiable function
 \begin{equation}
     \theta(x,\varepsilon) = 1 - e^{-x/\varepsilon}
 \end{equation} with $\varepsilon > 0$, noting that $\lim_{\varepsilon\to0} \theta(x,\varepsilon) = H(x)$. Defining 
 \begin{equation}
     \Theta(\mathbb{X},\varepsilon) = \sum_i \theta(\sigma_i(\mathbb{X}),\varepsilon)
 \end{equation}
 the goal is to solve Eq. (\ref{opt_pen}) with $\mathcal{R}(\mathbb{X}) = \Theta(\mathbb{X},\varepsilon)$. The non-differentiability of the rank function has been removed, but $\Theta(\mathbb{X},\varepsilon)$ is still concave.

A concave function has a special property: it is bounded \emph{above} everywhere by any tangent plane intersecting the function at one point. In other words, the value of the function at a given point (matrix), plus a term given by the inner product gradient of the function and the difference between the two points, will always be an upper bound to objective function itself. In this case the approximate rank function $\Theta$ has (elementwise) gradient
\begin{equation}
    \nabla \Theta(\mathbb{X},\varepsilon) = \frac{1}{\varepsilon}U
    \begin{pmatrix}
    e^{-\sigma_1(\mathbb{X})/\varepsilon} & & \\
    & e^{-\sigma_2(\mathbb{X})/\varepsilon} & \\
    & & \ddots
    \end{pmatrix} U^{\dagger},
\end{equation}
where $U\Sigma U^\dagger$ is the singular value decomposition of $\mathbb{X}$ (equivalent to its eigendecomposition, since $\mathbb{X}$ is positive semidefinite). Then, for any feasible matrix $\mathbb{X}_k$ obtained at the $k^{th}$ iteration,
\begin{equation}
    \Theta(\mathbb{X},\varepsilon) \leq \Theta(\mathbb{X}_k,\varepsilon) + \langle \nabla \Theta(\mathbb{X}_k,\varepsilon), \, \mathbb{X} - \mathbb{X}_k \rangle
    \label{eq:ThetaUB}
\end{equation}
The first term on the right-hand side of \eqref{ThetaUB} is constant with respect to $X$, while the second term is \emph{linear}. Hence minimizing the second term is a convex optimization problem, and will result in a minimization of \eqref{ThetaUB}, which is itself an upper bound on rank. We define the objective to minimize as $\mathcal{R}$:
\begin{equation}
    \mathcal{R}(\mathbb{X}) = \langle \nabla \Theta(\mathbb{X}_k,\varepsilon), \, \mathbb{X} \rangle
\end{equation}

Thus by iteratively solving convex optimization problems, we can continuously push down an upper bound on the rank, and thereby the rank itself, ideally near 1. This is an example of a ``majorization--minimization'' algorithm~\cite{Hunter2004}. To select an initial point for the optimization, we solve Eq. (\ref{opt_pen}) with $\mathcal{R}(\mathbb{X}) = \operatorname{Tr}(\mathbb{X})$.  For positive-semidefinite matrices, the trace is a well-known proxy for $\operatorname{rank}(\mathbb{X})$. Additionally, since $\Theta(\mathbb{X},\varepsilon)$ becomes highly concave as $\varepsilon \to 0$, which is liable to cause computational difficulty, we execute the MM algorithm iteratively over decreasing values of $\varepsilon$ until convergence.  Putting everything together, with convergence tolerances $\delta_1,\delta_2,\delta_3$, we implement the following process:
\begin{enumerate}
    \item Define arbitrary initial $\gamma$.
    \item Solve Eq. (\ref{opt_pen}) with $\mathcal{R}(\mathbb{X}) = \operatorname{Tr}(\mathbb{X})$ to obtain an initial $\mathbb{X}_k$.
    \item Define arbitrary initial $\varepsilon$.
    \item Solve Eq. (\ref{opt_pen}) with $\mathcal{R}(\mathbb{X}) = \langle \nabla \Theta(\mathbb{X}_k,\varepsilon), \, \mathbb{X} \rangle$ for the optimal solution $\mathbb{X} = \mathbb{X}_{k+1}$. Repeat until ${\lVert \mathbb{X}_{k+1} - \mathbb{X}_k \rVert_F}/{\lVert \mathbb{X}_k \rVert_F} < \delta_1$.
    \item Define the output of Step 4 as $\mathbb{X}_l$. Decrease $\varepsilon$ by an arbitrary factor $\alpha$ and repeat Step 4. Repeat until ${\lVert \mathbb{X}_{l+1} - \mathbb{X}_l \rVert_l}/{\lVert \mathbb{X}_l \rVert_F} < \delta_2$.
    \item Repeat steps 2-5, increasing $\gamma$ by an arbitrary factor $\beta$, until $\sigma_1(\mathbb{X})/\sigma_2(\mathbb{X}) > \delta_3$.
\end{enumerate}
We make simple (non-optimized) choices $\gamma = 10^{-7}$, $\delta_1 = 10^{-3}$, $\delta_2 = 10^{-3}$, $\delta_3 = 10^{5}$, $\epsilon = 0.5$, $\alpha = 2$, and $\beta = 1.5$.

\section{Metalens bound calculation}

\subsection{Formulation of the target function}

We consider a two-dimensional scattering problem for a TE-polarized electric field (such that it can be treated as a scalar electric field) and we seek to optimize the intensity of the field at a target point $x_\text{tar}$. 
We denote the field at the target point as $E_\text{tar} = E(x_\text{tar})$, and calculate a bound on the maximal value of the quantity $|E_\text{tar}|^2$, given a narrow design region of length $d$ in which each point in can be vacuum or a material with susceptibility $\chi_1$ (see Fig. 2(a) of the main text).
We can express the field at the target point as
\begin{align}
    E_\text{tar} = E_i(x_\text{tar}) + E_s(x_\text{tar}),
    \label{eq:Etar}
\end{align}
where $E_i(x_\text{tar})$ is the incident field at the target point, i.e., in the absence of material, and $E_s(x_\text{tar})$ is the scattered field at the target point.
Note, that the background field solves the background-$\LL$ operator, i.e., ${\LL_0 E_i =i\omega J}$.
The target point is set at a distance $f$ from the design region, such that the numerical aperture (given by ${\text{NA}=\sin[\arctan[(d/2)/f]]}$) is set to $\text{NA}=0.9$.

We can express the scattered field at the target point using a Green's function 
\begin{align}
{E_s(x_\text{tar}) = \int \Gamma_0(x_\text{tar},x) P(x)}dx,
\label{eq:Es_GF}
\end{align}
where $P(x)$ is the polarization currents in the design material, given by 
\begin{align}
P(x)=\chi(x)E(x)=\frac{1}{\omega^2}(\LL_0 E - i\omega J(x))=\frac{1}{\omega^2}(\LL_0 E - \LL_0 E_i).
\label{eq:P}
\end{align}
For clarity of notation, we will denote the integration as an operator ${\Gamma_\text{tar} = \omega^{-2}\int  \Gamma_0\left(x_\text{tar},x\right)dx}$ (numerically, this a row from the matrix representing the Greens' function).

A natural way to proceed is to substitute Eqs. (\ref{eq:Es_GF}) and (\ref{eq:P}) into Eq. (\ref{eq:Etar}) and squaring:
\begin{align}
    |E_\text{tar}|^2 &= \left|E_i\left(x_\text{tar}\right) + \Gamma_\text{tar} \left( \LL_0 E - \LL_0 E_i\right) \right|^2\\ 
    &=\left|E_i\left(x_\text{tar}\right)\right|^2 + \left|\Gamma_\text{tar} \left(\LL_0 E - \LL_0 E_i\right) \right|^2 + 2 \text{Re} \left[ E_i^\dagger\left(x_\text{tar}\right)\Gamma_\text{tar} \left( \LL_0 E - \LL_0 E_i\right) \right].
\end{align}
However, in this case, the target function has a matrix of the form $\Gamma_\text{tar}^\dagger \Gamma_\text{tar}$, which is not sparse, and we will not be able to exploit the sparse structure of the problem in the optimization.

An alternative formulation of the target function, which preserves sparsity, is given by
\begin{align}
    |E_\text{tar}| = \max_\theta \ \text{Re}\left[ E_\text{tar} e^{i\theta} \right],
    \label{eq:theta_Etar}
\end{align}
where $\theta$ is an auxiliary angle. After the optimization procedure, we can square the result to obtain $|E_\text{tar}|^2$.
Explicitly, we have 
\begin{align}
    |E_\text{tar}| =\max_\theta \  \text{Re}\left[E_i\left(x_\text{tar}\right)e^{i\theta} + \Gamma_\text{tar}\left(x_\text{tar},x\right) \left( \LL_0 E - \LL_0 E_i\right)e^{i\theta} \right],
    \label{eq:target_fun_E}
\end{align}
which is linear in the term $\Gamma_\text{tar}$ and preserves the sparse nature of the problem.
Together with Eq. (\ref{eq:constEM}), this forms the QCQP to be solved.

\begin{figure*}[htb]
    \centering
    \includegraphics[scale=.6]{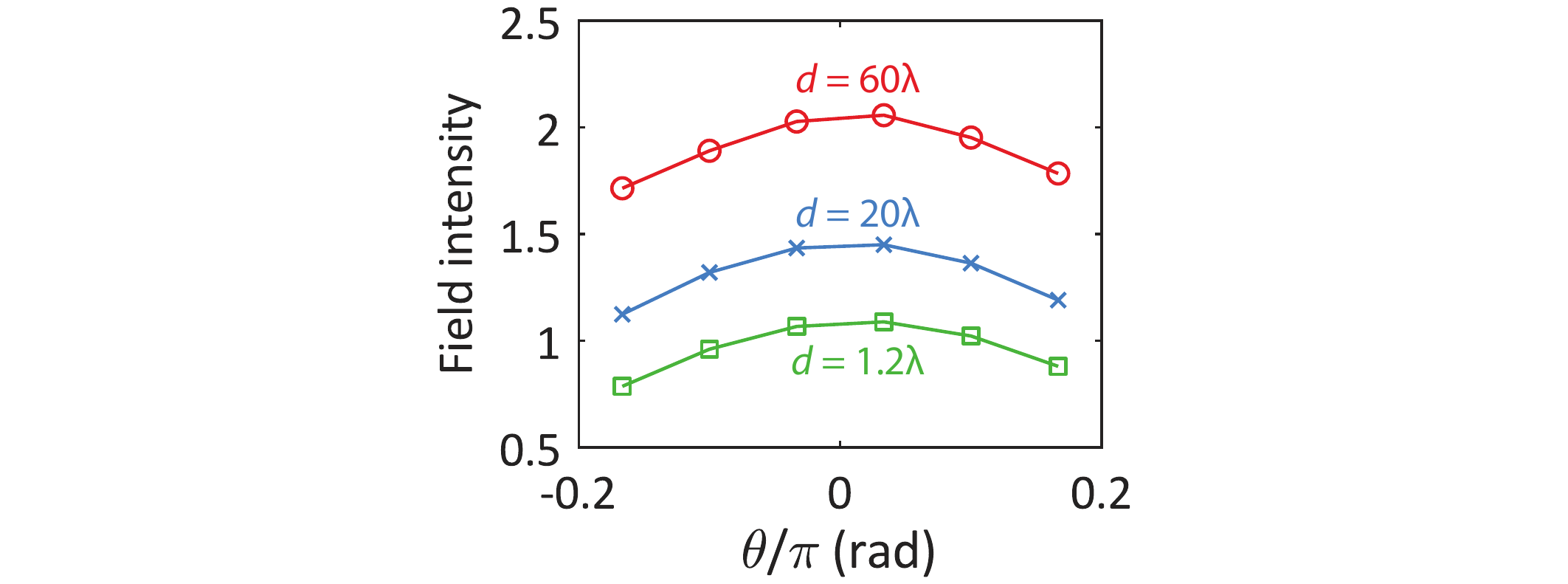}
    \caption{Optimization results for the field intensity focused by a metalens of different lengths ($d$), using an auxiliary angle $\theta$ (see Eq. (\ref{eq:theta_Etar})).
    }
    \label{fig:theta_sweep}
\end{figure*}

Fig. \ref{fig:theta_sweep} shows the results of the optimization process for different values of $\theta$, from which the maximum can be selected. While this approach requires multiple optimizations (sweeping over different values of $\theta$), the speed-up provided by the sparseness of this form is significantly faster compared to non-sparse formulations, and enables the optimization over large domains (see Fig. 2(d) in the main text).

\subsection{Scattered-field formulation}

The target function in Eq. (\ref{eq:target_fun_E}), as well as the constraints from Eq. (\ref{eq:constEM}) are quadratic form in the total field (E). Alternatively, the QCQP can be written where the quadratic form is of the scattered field $E_s$ (given by $E_s = E - E_i$).
Using Eqs. (\ref{eq:Es_GF}) and (\ref{eq:P}) we can rewrite the target function in terms of $E_s$:
\begin{align}
    |E_\text{tar}|^2 = \left|E_i\left(x_\text{tar}\right) + \Gamma_\text{tar} \LL_0 E_s \right|^2.
\label{eq:target_Es}    
\end{align}

Similarly, we can express the constraints from Eq. (\ref{eq:constEM}) in terms of scattered field:
\begin{align}
    &\braket{\left( \LL_0 \left(E_i + E_s\right) - i\omega{J}\right) , \frac{\chi_c^{-1}}{\omega^2}\left( \LL_0 \left(E_i + E_s\right) -i\omega{J}\right)}_{v_i} \\
    &\qquad\qquad - \braket{ \LL_0 \left(E_i + E_s\right) , \left(E_i + E_s\right)}_{v_i} -i\omega \braket{J , \left(E_i + E_s\right)}_{v_i} =0,\quad {i=1,2,3\dots}.
\end{align}
Using $i\omega J=\LL_0E_i$ and rearranging, we are left with:
\begin{align}
    \braket{ \LL_0  E_s  , \frac{\chi_c^{-1}}{\omega^2} \LL_0  E_s}_{v_i} - \braket{ \LL_0  E_s , E_s}_{v_i} - \braket{ \LL_0  E_s , E_i}_{v_i} =0,\quad {i=1,2,3\dots}.
    \label{eq:const_Es}
\end{align}
Together, the target function (Eq. (\ref{eq:target_Es})) and the constraints (Eq. (\ref{eq:const_Es})) form a QCQP in the scattered field $E_s$.

\subsection{Unitary bound}

In this section, we derive a bound for the efficiency of a metalens of length $d$ based on the unitarity of the scattering matrix describing the system.
In the most general case, we can write the field at the target point using Green's functions:
\begin{align}
    E_\text{tar}=\int_0^d \Gamma_{zz}^{EE}(x_t,x) \xi_z^E(x) dx +\int_0^d \Gamma_{zy}^{EH}(x_t,x) \xi_y^H(x) dx,
\end{align}
where $\Gamma_{zz}^{EE}$ and $\Gamma_{zy}^{EH}$ are the electric and magnetic Green's functions, respectively, and $\xi_z^E$ and $\xi_y^H$ represent electric and magnetic current sources, respectively. 
After discretizing the domain, we have the matrix form:
\begin{align}
    E_\text{tar}= \underbrace{\begin{pmatrix} \Gamma_{zz}^{EE} & \Gamma_{zy}^{EH}\end{pmatrix}}_{\Gamma} \underbrace{\begin{pmatrix} \xi_z^E \\ \xi_y^H\end{pmatrix}}_{\xi} h,
\end{align}
where $h$ is the discretization step size, and squaring the field at the target point, leaves us with:
\begin{align}
    \left|E_\text{tar}\right|^2= \xi^\dagger\Gamma^\dagger\Gamma\xi h^2.
\end{align}

Assuming the incident field is a plane wave with amplitude 1, we can normalize the current distributions:
\begin{align}
    \xi^\dagger\xi h = \int \left| n \times E\right|^2 dx + \int \left| n \times H \right|^2 dx = 2d.
\end{align}

We can now write the intensity maximization problem:
\begin{align}
\begin{split}
&\max \quad \xi^\dagger\Gamma^\dagger\Gamma\xi h^2\\ 
&\text{s.t.} \qquad \xi^\dagger \xi = \frac{2d}{h}
\end{split}
\end{align}
In order to solve the equation above, we define a normalized current $\zeta=\xi\sqrt{h/(2d)}$, such that the problem can be rewritten in the following form:
\begin{align}
\begin{split}
&\max \quad \zeta^\dagger \left(2d\Gamma^\dagger\Gamma h\right)\zeta\\ 
&\text{s.t.} \qquad \zeta^\dagger \zeta = 1
\end{split}
\end{align}
The optimization problem is now in the form of a Rayleigh quotient, where the maximum of the target function is given by the largest eigenvalue of $2d\Gamma^\dagger\Gamma h$.
The results of this bound calculation are shown in Fig. 2(c) of the main text.

\begin{figure}[h]
    \centering
    \includegraphics[scale=.70]{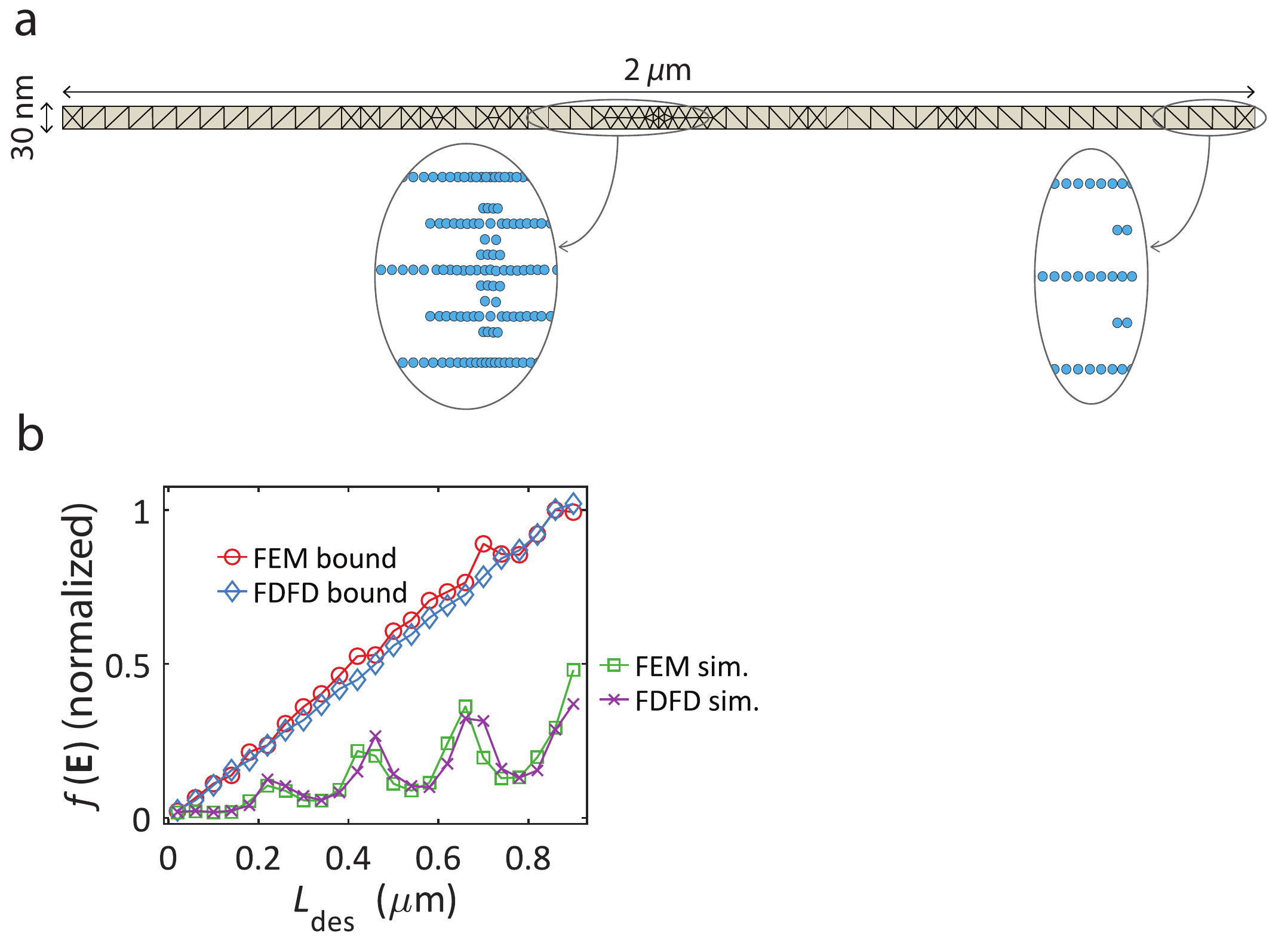}
    \caption{
    \textbf{(a)} COMSOL-generated mesh, and a magnified view of the degrees of freedom being solved for. 
    % The orange points represent the design region (this is an example for a specific length $L_\text{des})$.
    \textbf{(b)} Calculated bound as a function of design region length, using differential operators from FEM (red) and FDFD (blue) solvers.
    Simulation results of an unstructured dielectric (using both solvers) are shown for reference.
    }
    \label{fig:COMSOL}
\end{figure}

\section{Compatibility with FEM solvers}
The formulation we present here utilizes the differential form of the physical equation, which is used in many commercially available solvers.
These can be both finite difference (FDFD) solvers, as well as finite element (FEM) solvers.
Here, we demonstrate how an FEM electromagnetic solver (COMSOL) can be used to calculate a bound in a simple photonic problem.
We model a 2D system, with a geometry shown in Fig. \ref{fig:COMSOL}(a).
We set periodic boundary conditions for the top and bottom boundaries and simulate perfectly matched layers (PML) on the sides on the structure.
Fig. \ref{fig:COMSOL}
After setting current sources, we export the matrix and vector describing the linear system as constructed by the software.
We choose the point that comprise a `design' region (see Fig. \ref{fig:COMSOL}(b) for an example) and calculate bounds for the maximal value of the target function
\begin{align}
    f(\textbf{E})=\int_{V_\text{des}}|\textbf{E}(\textbf{x})|^2 d\textbf{x},
\end{align}
i.e., integrated field intensity over the entire design region $V_\text{des}$.
In this example, the material has a refractive index of $n=\sqrt{12}$, and the optical wavelength was set to $\lambda=1.55 \mu\text{m}$.
For comparison, we repeat the calculation with a finite-time frequency domain solver (FDFD).
Fig. \ref{fig:COMSOL}(c) shows the calculated bounds as a function of design region length.
We see that the FDFD and FEM based bound calculations agree, showing a linear dependence of the target function with the design region length.

\section{Discussion of related work}
\citeasnoun{Angeris2021} is a delightful review of approaches to inverse design and performance bounds, especially techniques that had been developed by early 2021. Yet \citeasnoun{Angeris2021} goes significantly further, offering a unified, generalized mathematical framework for these approaches. Here we discuss the bound approach, newly developed in \citeasnoun{Angeris2021}, that is close to the approach we propose in the main text.

The starting point of \citeasnoun{Angeris2021} is a physical design problem written in the form 
\begin{align}
    \left(A_0 + \diag\left(\theta\right)\right)z = b,
    \label{eq:physdesign}
\end{align}
where $A_0$ is a design-independent differential operator (plus, possibly, the identity matrix scaled by background constants), $\diag(\theta)$ is a diagonal matrix whose diagonal elements are the designable degrees of freedom, $z$ is the field of interest, and $b$ comprises the sources. 

The matrices and vectors in the physical design equation are all assumed to be real-valued. From this assumption, they derive ``power inequalities'' of the form
\begin{align}
    \left(A_0 z - b\right)^T D \left(A_0 z - b\right) \leq z^T D z,
\end{align}
which is equivalent to the power-conservation bounds of \cite{Kuang2020,Molesky2020}, when all of the equations are real-valued.

Yet in nearly all cases of interest, the fields and design variables \emph{cannot} be trivially assumed as real-valued. Complex values are of course built into quantum dynamics. In electromagnetism, open boundaries require complex-valued fields (e.g., through perfectly matched layers). Plasmonic structures require loss. Perhaps the one exception is the calculation of modes of closed, periodic systems (e.g. photonic crystals), for eigenstate properties. For any scattering problem with a source, however, the power from the source must be dissipated in either the material or into far-field radiation, with complex-valued fields as a result. 

The formulation of \citeasnoun{Angeris2021} does not lead to a design QCQP when the fields are complex-valued. In order to transform the original, complex-valued physical design equation to a real-valued form of the type of \eqref{physdesign}, one must perform a standard doubling of the dimensions of the vectors and matrices, separating the real and imaginary parts of all variables involved. (See \citeasnoun{angeris2022bounds} for detailed descriptions of these steps.) This leads to a vector $\theta$ of design variables that is twice as large as the number of degrees of freedom, and within which the first and second halves of the vector should be equal. Such a constraint is a pain to enforce, however (it possibly could be enforced by the methods of \citeasnoun{Shim2021}), and instead is dropped in \cite{Angeris2021,angeris2022bounds}. Hence, that approach leads to a relaxed QCQP that can be used for bounds, but \emph{not} a QCQP formulation of the design problem, as proposed in the main text.

\bibliographystyle{unsrtnat}
\bibliography{diff_bounds}